\documentclass[review,authoryear]{elsarticle}
\usepackage{amssymb}
\usepackage{amsmath}

\usepackage{caption}
\usepackage{subcaption}
\usepackage{changepage}
\usepackage{booktabs}

\def\hh{\mskip 1mu}

\journal{Icarus}
\begin{document}

\begin{frontmatter}

\title{Vertical structures induced by embedded moonlets in Saturn's rings: the gap region}

\author{H. Hoffmann\corref{cor1}}
\cortext[cor1]{Corresponding author.}
\ead{hohoff@uni-potsdam.de}
\author{M. Sei\ss}
\author{F. Spahn}
\address{Nonlinear Dynamics Group, Institute of Physics and Astronomy, University of Potsdam, 14476 Golm, Germany}

\begin{abstract}
We study the vertical extent of propeller structures in Saturn's rings. Our focus lies on
the gap region of the propeller and on non-inclined propeller moonlets. In order to describe
the vertical structure of propellers we extend the model of \citet{spahn00} to include the
vertical direction. We find that the gravitational interaction of ring particles with the
non-inclined moonlet does not induce considerable vertical excursions of ring particles, but
causes a considerable thermal motion in the ring plane. We expect ring particle collisions
to partly convert the lateral induced thermal motion into vertical excursions of ring particles.
For the gap region of the propeller, we calculate gap averaged propeller heights on the order
of 0.7 Hill radii, which is of the order of the moonlet radius. In our model the propeller
height decreases exponentially until viscous heating and collisional cooling balance. We
estimate Hill radii of $370$m and $615$m for the propellers Earhart and Bl\'eriot. Our model
predicts about $120$km for the azimuthal extent of the Earhart propeller at Saturn's 2009 equinox,
being consistent with values determined from Cassini images.
\end{abstract}


\end{frontmatter}

\section{Introduction}
\label{sec:introduction}

Planetary rings are one of the most remarkable and beautiful cosmic structures. They are
considered to be natural dynamical laboratories \citep{burns06}, exemplifying the physics
of cosmic disks, such as accretion disks or galactic disks, which are much larger and much
farther away from Earth. An exciting example is the presence of small moons embedded in
Saturn's rings, henceforth called moonlets, which have their analog in planetary embryos
orbiting within a protoplanetary disk \citep{artymowicz06,papaloizou07}.

The fact that the resolution of even \emph{Cassini's} cameras is too
low to image these moonlets directly, brought up the idea of investigating moonlet induced
putative structures in the rings \citep{spahn87,spahn89}, with the hope that these features
could be captured by the spacecraft cameras or instruments. This then led to predictions
of the propeller structures \citep{spahn00,sremcevic02} which are carved in the rings by
the moonlet. Subsequent numerical particle experiments \citep{seiss05,sremcevic07,lewis09}
completed the \emph{fingerprint} of such gravitational perturbers and confirmed the spatial
scaling of the propeller structure. Depending on its size, an embedded ring-moon either
creates a propeller (sizes below 1 km) or, alternatively, opens up a circumferential gap
(for sizes $>$ 1 km, e.g. the ring-moons {\em Pan} and {\em Daphnis}). Both structures,
propeller and circumferential gap, are decorated with density wakes, completing the
structural picture.  Up to this stage, all these density features have been assumed to
occur only in the ring plane, a vertical stratification of moonlet induced structures has
not seemed to be of importance.

More than 150 propeller moonlets have now been detected \citep{tiscareno06,tiscareno08}
and among them a few which are large enough to allow \emph{Cassini's} cameras to take
several snapshots of their propellers at different times, confirming in this way their
orbital motion. Those moonlets were nicknamed after famous aviators, e.g.: Bl\'{e}riot,
Kingsford Smith, Earhart \citep{tiscareno10}.

In the summer of 2009, at Saturn's equinox (the Sunset at Saturn's rings), the perfect
opportunity arose to detect any vertical structure deviating from the mean ring plane
by observing shadows cast on the rings. At this time the density structures around the
largest propeller moonlets, as well as those around the ring-moon \emph{Daphnis}, created
prominent shadows. These can be assigned to the wakes and in the case of the propeller
moonlets also to excited regions of the propeller, where the moonlet induces a partial
gap. The shadows were much longer than the moon's size itself. The lateral extent of
the shadows allows to conclude that moonlet induced vertical excursions of ring particles
can be in the range of several kilometers in the case of \emph{Daphnis} or several hundred
meters in the case of the large propeller moonlets. These very facts directly indicate
the necessity to investigate the vertical stratification of moonlet induced structures,
which has not been the focus of former models of the moonlet's \emph{fingerprint}.
In this work we will study the vertical extent of moonlet induced propeller stuctures,
focusing on the gap region of the propeller.

The paper is organized as follows: In Section \ref{sec:extended_model} the extended propeller
model is presented. In Section \ref{sec:gravitational_scattering} the mass flow through the
scattering region is calculated by a probabilistic approach, and values of the moonlet induced
thermal velocities are determined, which are later used as initial conditions for the hydrdynamical
equations. Section \ref{sec:hydrodynamic_flow} gives the hydrodynamical balance equations,
which we use to model the diffusion of mass into the induced gap and the relaxation of the
ring temperature. In Section \ref{sec:propeller_height} we calculate the height of the ring
in the gap region of a propeller. Section \ref{sec:discussion} discusses made assumptions
and the application of our results to propeller features in Saturn's rings. Our results are
summarized in Section \ref{sec:conclusions}.

\section{Extended model of gravitational scattering}
\label{sec:extended_model}

\subsection{The scattering region}
\label{sec:scattering_region}

The first step in the formulation of our model, is to divide the planetary
ring, composed of granular grains and one moonlet, into two regions:

\begin{enumerate}
\item[(i)] the scattering region,
\item[(ii)] the rest of the ring.
\end{enumerate}

In this work we consider moonlets on circular, i.e. on non-inclined and non-eccentric,
orbits. The scattering region is the small area (volume) around the moonlet where the
majority of the trajectory changes, due to the moonlet's gravity, take place. This region
of the embedded moonlet's gravitational influence is of the order of a few Hill radii
\begin{equation}
h = a_0\ \left( \frac{m_m}{3(m_m + m_s)}\right)^{1/3}\ ,
\label{eq:hill}
\end{equation}
where $a_0$ is the semimajor axis of the moonlet, $m_m$ its mass and $m_s$ the mass of
Saturn. Compared to the moonlet's semimajor axis the Hill radius is usually very small.
For large propeller moonlets, like Bl\'{e}riot or Earhart, the ratio $h^* = h/a_0$ is
approximately $10^{-6}$ \citep{tiscareno10}. This low ratio naturally allows the splitting
of the rings into the two regions, where the scattering region is shrunk to a line.

For the rest of the ring, where the moonlet's gravity is negligible, the moonlet induced
structures are assumed to relax due to inelastic collisional cooling and viscous diffusion
\citep{spahn00, sremcevic02}. In previous hydrodynamic models the former process, the
relaxation of the granular temperature $T \sim c^2/3$, has not been considered. This thermal
relaxation and 3D-scattering are the major themes of this work.

%

\subsection{Encounter with the moonlet -- gravitational scattering}
\label{sec:gravitational_scattering}

We describe the encounter of ring particles with the moonlet in a corotating
frame, rotating about Saturn with the mean motion of the moonlet. The frame's
origin coincides with the mean orbital location of the moonlet. The x axis points radially
outward, the y axis points into the azimuthal direction and the z axis is normal to
the ring plane in a way that the axes form a right handed coordinate system.

With this, the dynamics of ring particles in the corotating frame is given by
\begin{equation}
\label{eq:motion}
\ddot{\mathbf{r}} + 2\:\!\mathbf{\Omega}_0 \times \dot{\mathbf{r}}
   + \mathbf{\Omega}_0 \times (\mathbf{\Omega}_0 \times \mathbf{r})
   = -\mathbf{\nabla}\Phi_s - \mathbf{\nabla}\Phi_m,
\end{equation}
where $\Phi_s$ and $\Phi_m$ are the gravitational potentials due to Saturn and the moonlet.

The ring particles are on orbits with low eccentricity and inclination and the mass
of the moonlet is very small compared to Saturn's mass $m_m/m_s \ll 1$. In the
vicinity of the moonlet the equations of motion of ring particles are well approximated
by Hill's equations \citep{hill78,henon86}.

With the scaled coordinates $\tilde{x} = x/h$, $\tilde{y} = y/h$, $\tilde{z} = z/h$
and scaled time $t' = \Omega_0 t$ Hill's equations become
\begin{align}
\ddot{\tilde{x}} &= \phantom{-}2 \dot{\tilde{y}} + 3 \tilde{x} - 3\tilde{x}/\tilde{s}^3 \nonumber \\
\ddot{\tilde{y}} &= -2 \dot{\tilde{x}} - 3\tilde{y}/\tilde{s}^3\label{eq:hill_eq} \\
\ddot{\tilde{z}} &= -\tilde{z} - 3\tilde{z}/\tilde{s}^3 \nonumber \ ,
\end{align}
where $\tilde{s}^2 = \tilde{x}^2 + \tilde{y}^2 + \tilde{z}^2$ is the scaled distance
to the moonlet and $\dot{\tilde{x}} = d\tilde{x}/dt'$. These equations are point symmetric
about the position of the moonlet ($\tilde{x} = \tilde{y} = \tilde{z} = 0$), and quite
comfortable, they do not depend on the moonlet mass anymore. All information of the moonlet
mass is contained in the scaling length $h$.

When the ring particles are not in the vicinity of the moonlet, i.e. $1/\tilde{s}^3\to 0$ and 
therefore $|\mathbf{\nabla}\Phi_s|\to 0$, their trajectories are well described by the solutions
to the homogeneous Hill's equations
\begin{align}
\tilde{x}(t') &= \tilde{a} - \tilde{e} \cos(t' + \tilde\tau) \nonumber \\
\tilde{y}(t') &= C - \frac{3}{2} \tilde{a} t' + 2\tilde{e} \sin(t' + \tilde\tau)\label{eq:homhill}\\
\tilde{z}(t') &= \tilde{\imath} \sin(t' + \tilde\omega) \ .\nonumber
\end{align}
The semimajor axis, eccentricity and inclination are scaled according to
\begin{equation}
\tilde{a} = \frac{a - a_0}{h}, \quad
\tilde{e} = \frac{e\:\!a_0}{h}, \quad
\tilde{\imath} = \frac{i\:\!a_0}{h}\ ,
\end{equation}
where a spherically symmetric planet ($\Phi_s \sim -G m_s / r$) is assumed. The phases $\tilde\tau$ and
$\tilde\omega$ are the longitude of pericenter and the longitude of the ascending node respectively.

\subsubsection{Test particle integrations}
\label{sec:test_particle_integrations}

We integrate the equations of motion numerically for a set of test particles using a 5th
order embedded Runge-Kutta scheme with adaptive step size control \citep{press92}. The
moonlet and Saturn are assumed to be point masses. The integrations start upstream of the
moonlet at an azimuthal distance of 1000 Hill radii to the moonlet and are terminated when
the test particle's azimuthal distance exceeds 1000 Hill radii. These limits ensure that
the region of interaction is well included in the integration.

We take the initial semimajor axis of the test particles to be uniformly distributed in
the range of $-20$ to $20$ Hill radii radial distance to the moonlet. The initial eccentricities
and inclinations of the test particles are choosen to be Rayleigh distributed 
\begin{align}
f(\tilde{e}, \tilde{\imath}) =
    \frac{\tilde{e}}{\tilde{c}_x^{\:\!2}} \exp\left( - \frac{\tilde{e}^{\:\!2}}{2\tilde{c}_x^{\:\!2}} \right)
    \frac{\tilde{\imath}}{\tilde{c}_z^{\:\!2}} \exp\left( - \frac{\tilde{\imath}^{\:\!2}}{2\tilde{c}_z^{\:\!2}} \right) \quad,
\end{align}
with uniformly distributed initial phases $\psi$ and $\zeta$. This is a fair assumption for
many kinds of disks \citep{petit87,ida92,lissauer93,ohtsuki00}.

This distribution of orbital elements corresponds to a triaxial Gaussian velocity distribution
\begin{equation}
\label{eq:velodistro}
f(\tilde{v}_x, \tilde{v}_y, \tilde{v}_z) =
    \frac{1}{\pi\tilde{c}_x^{\hh 2}} \exp\left( - \frac{\tilde{v}_x^2 + 4\tilde{v}_y^2}{\tilde{c}_x^{\hh 2}} \right)\ 
    \frac{1}{\sqrt{2\pi\tilde{c}_z^{\hh 2}}} \exp\left( -\frac{\tilde{v}_z^2}{\tilde{c}_z^{\:\!2}} \right)
\end{equation}
with a diagonal velocity dispersion tensor $\textbf{T}$, where $T_{xx} = \tilde{c}_x^{\,2}$,
$T_{yy} = (\tilde{c}_x/2)^2$ and $T_{zz} = \tilde{c}_z^{\,2}$. The scaled quantities $\tilde{c}_x$,
$\tilde{c}_z$ are related to the unscaled ones by
\begin{equation}
\tilde{c}_x = c_x / (\Omega_0 h),\qquad
\tilde{c}_z = c_z / (\Omega_0 h).
\end{equation}

We choose the ratio $\tilde{c}_z/\tilde{c}_x = 0.65$, which is consistent
with the above Gaussian velocity distribution \citep{goldreich78} and which is a reasonable
value if we neglect self gravity \citep{salo01}.

This deviation from the Maxwellian velocity distribution of conservative gases is an expression
of the non-equilibrium state, caused by the steady dissipation of orbital energy in collisions.
This energy loss is counterbalanced by the energy gain due to viscous Keplerian shear, guaranteeing
a quasi-equilibrium manifested by a steady state. This energy pumping goes at the expense of the
collective Keplerian motion and thus the ring particles spiral slowly into Saturn in the long
term ($10^8$ years).

During the trajectory integrations, the test particles and the moonlet are assumed
to be point masses and the minimal distance of the test particles to the moonlet is
recorded. For simplicity, we ignored particles which would collide with a spherical
moonlet of finite size in further calculations. To estimate the difference, we also
examined the case where all particles are used. Although the radial position of the
maximally induced ring temperature changes a bit, we found no significant changes in
the resulting gap averaged heights.

\subsubsection{Mass transfer}
\label{sec:mass_transfer}

We use the approach of \citet{spahn89} to describe the mass transfer through the
scattering region. Motivated by the chaotic behaviour of single particle trajectories
near the moonlet \citep{petit86}, the encounter of ring particles with the moonlet
is modelled by a probabilistic Markov chain model. The results of the test particle
integrations are used to calculate transitional probabilities between initial ($\tilde{x}$, $\tilde{z}$)
and final ($\tilde{x}'$,$\tilde{z}'$) positions of the test particles at the azimuthal boundary of
the scattering region. These probabilities define a scattering operator $A$,
where $A(\tilde{x}',\tilde{z}'\,|\,\tilde{x},\tilde{z})\,d\tilde{x}\,d\tilde{z}$ is
the probability that matter is scattered from
$(\tilde{x},\tilde{x}+d\tilde{x}) \times (\tilde{z},\tilde{z}+d\tilde{z})$
to $(\tilde{x}',\tilde{x}'+d\tilde{x}') \times (\tilde{z}',\tilde{z}'+d\tilde{z}')$.
The primes denote values after the scattering.

We assume that the scattering region can be approximated by the $\tilde{x}$-$\tilde{z}$ plane
at $\tilde{y} = 0$, which connects Saturn and the moonlet
and is analogous to the scattering line of \citet{spahn89}. In our model, the complete
action of the moonlet on the ring particles happens at this scattering plane. The
scattering operator relates the azimuthal mass flux entering the scattering region
to the azimuthal mass flux leaving the scattering region
\begin{equation}
\left|J_{\tilde{y}}(\tilde{x}',\tilde{y}=0^\pm,\tilde{z}')\right| =
   \iint\,d\tilde{x}\,d\tilde{z}\ 
   A(\tilde{x}',\tilde{z}'\,|\,\tilde{x},\tilde{z})\,\left|J_{\tilde{y}}(\tilde{x},\tilde{z},\tilde{y}=0^\mp)\right|\ .
\end{equation}

Mass conservation is expressed as a condition for the scattering operator
\begin{equation}
\iint\,d\tilde{x}'\,d\tilde{z}'\ A(\tilde{x}',\tilde{z}'\,|\,\tilde{x},\tilde{z}) = 1\ ,
\end{equation}
also illustrating that accreting particles are ignored here.

For the numerical calculation of the scattering operator we divide the radial and the
vertical direction into bins. The bin $(j,k)$ describes the region
$(\tilde{x}_j, \tilde{x}_{j+1}) \times (\tilde{z}_k, \tilde{z}_{k+1})$, $j$ being an
index in radial direction and $k$ one in vertical direction. In the following,
primed bin indices refer to the situation after the scattering by the moonlet and unprimed
ones to the situation before.

The discretized scattering operator is then calculated by
\begin{equation}
\label{eq:scattop}
A(j',k'|j,k) = \frac{N(j',k',j,k)}{N(j,k)}\ ,
\end{equation}
where $N(j,k)$ is the number of test particles starting in bin $(j,k)$. We use an averaging
procedure to calculate the number of test particles $N(j',k',j,k)$ starting in bin $(j,k)$
and ending in bin $(j',k')$. For each radial end bin $j'$, describing the interval
$(\tilde{x}_{j'}, \tilde{x}_{j'+1})$, the time per orbit, the test particle will stay in
this radial end bin is calculated \citep{spahn87}
\begin{equation}
p(j') = \frac{\Delta t(j')}{T} =
   \frac{1}{\pi} \left|\, \arccos\left( \frac{\tilde{a} - \tilde{x}_{j'+1}}{\tilde{e}} \right)
   - \arccos\left( \frac{\tilde{a} - \tilde{x}_{j'}}{\tilde{e}} \right) \right|\ .
   \label{eq:orbit_averaging}
\end{equation}
$N(j',k',j,k)$ is then calculated by
\begin{equation}
N(j',k',j,k) = \sum_n p^{(n)}(j')\,\delta^{(n)}(k')\,\delta^{(n)}(j,k)\ ,
\end{equation}
where the sum is over all test particles and $\delta^{(n)}(j,k)$ is $1$ for test
particles starting in bin $(j,k)$ and $0$ otherwise.

%
%
%
Outside the scattering region we describe the ring using hydrodynamical equations.
We assume that the fluid parcels entering and leaving the scattering region have a
Keplerian azimuthal mean velocity $\tilde{u}_{\tilde{y}}(\tilde{x}) = -3\tilde{x}/2$.
The azimuthal mass flux at the scattering plane is then \citep{spahn89}
\begin{equation}
J_{\tilde{y}}(\tilde{x},\tilde{y}=0^\pm,\tilde{z}) = \rho(\tilde{x},\tilde{y}=0^\pm,\tilde{z})\ \tilde{u}_{\tilde{y}}(\tilde{x})\ ,
\end{equation}
where $\rho$ denotes the mass density. The mass transfer through the scattering region is thus calculated by
\begin{equation}
\rho(j',k') = \frac{1}{|\tilde{u}_{\tilde{y}}(j')|} \sum_{j,k} A(j',k'\,|\,j,k)\,
    \rho(j,k)\,\left| \tilde{u}_{\tilde{y}}(j) \right|\ .
\label{eq:dens_calc}
\end{equation}

\begin{figure}
\begin{adjustwidth}{-1in}{-1in}%
\begin{subfigure}[t]{0.7\textwidth}
    \centering
    \includegraphics[width=0.95\textwidth]{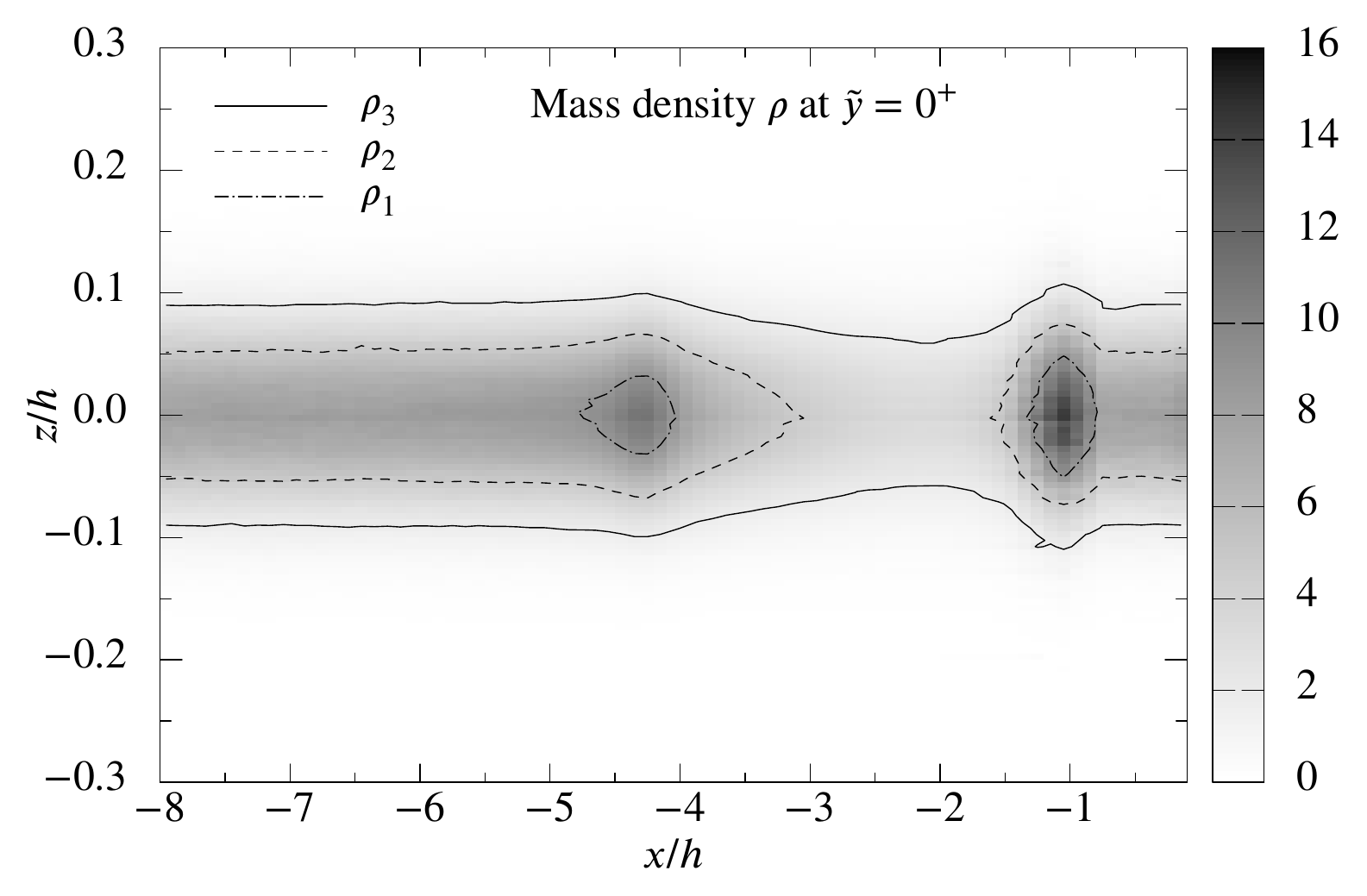}
    \caption{}\label{fig:density3D}
\end{subfigure}%
\qquad
\begin{subfigure}[t]{0.7\textwidth}
    \centering
    \includegraphics[width=\textwidth]{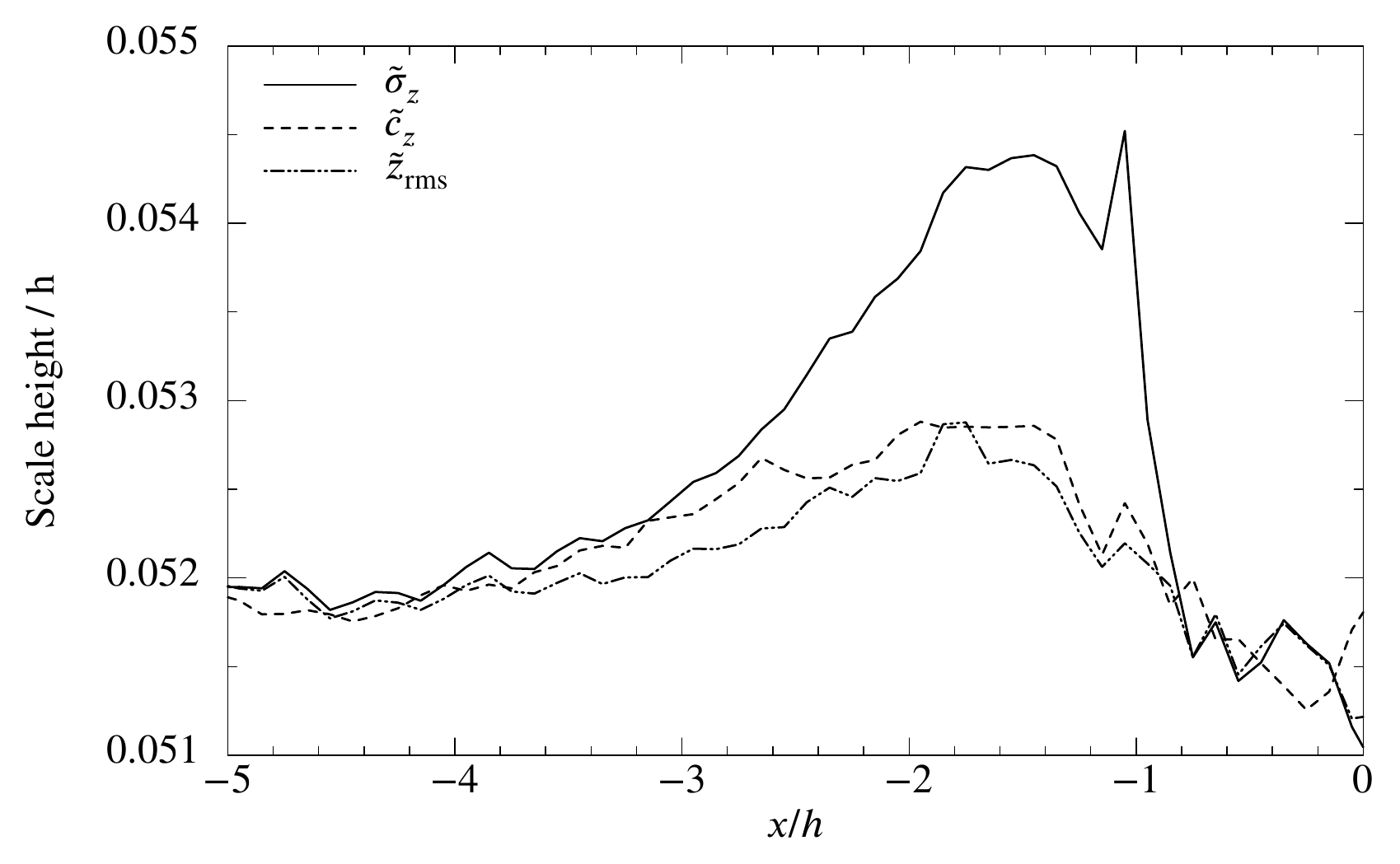}
    \caption{}\label{fig:zsq}
\end{subfigure}%
\caption{(a) Mass density $\rho$ scaled by $\Sigma_0/h$ after scattering by a moonlet with a Hill
radius of $300$m. The contour lines are for $\rho_1=1.15\,\rho_0(\tilde{z}=0)$, $\rho_2=\rho_0(\tilde{z}=\tilde{c}_z)$
and $\rho_3=\rho_0(\tilde{z}=\sqrt{3}\,\tilde{c}_z)$, where $\rho_0$, given by equation (\ref{eq:rho0}), is the mass
density before the scattering by the moonlet. Clearly seen is the less dense gap region between
$1.5 \le |\tilde{x}| \le 4$.  (b) Ring thickness after the scattering by the same moonlet measured by:
standard deviation $\tilde{\sigma}_z$ of the vertical mass density profile, root mean square
$\tilde{z}_\text{rms}$ of the vertical excursions and vertical thermal velocity $\tilde{c}_z$, all
at $\tilde{y}=0^+$. All three are within a few percent of the values before the scattering by the
moonlet, illustrating that the moonlet's gravity alone does not induce considerable vertical excursions
of ring particles.}\label{fig:massdensities}
\end{adjustwidth}
\end{figure}

We assume the ring to be in an equilibrium state before the encounter with the moonlet.
In the following all quantities of the unperturbed ring have the subscript $0$. The
surface mass density $\Sigma_0$ shall be constant, so that the mass density in the
thin-disk approximation is given by
\begin{equation}
\rho_0(\tilde{x}, \tilde{y}, \tilde{z}) = \frac{\Sigma_0/h}{\sqrt{2\pi\tilde{c}_{z0}^2}}\
   \exp\left(  - \frac{\tilde{z}^2}{2\tilde{c}^2_{z0}} \right)\ .
\label{eq:rho0}
\end{equation}
%

The two plots in Figure \ref{fig:massdensities} are made for a moonlet with $300$m Hill
radius and the orbital parameters of Earhart \citep{tiscareno10}. We have choosen the
velocity dispersion of the unperturbed ring to be $c_0 = 3.9$mm/s, corresponding to
$\tilde{c}_z = 0.052$. Because of the point symmetry of the Hill equations (\ref{eq:hill_eq}),
we show only values for $\tilde{x} \le 0$ in the plots.

Figure \ref{fig:density3D} shows the mass density $\rho$ scaled by $\Sigma_0/h$ after
the scattering by the moonlet. The mass density in the regions $|\tilde{x}| < 1$ and
$|\tilde{x}| > 5$ is nearly unchanged, whereas the gap region $1.5 < |\tilde{x}| < 4$
has a considerably lower mass density than the ring before the scattering by the moonlet.
There are two regions of high mass density around $|\tilde{x}| = 1$ and $|\tilde{x}| = 4.25$.
The solid line ($\phi=0.0$) in Figure \ref{fig:density_phi} shows the corresponding surface
mass density.

Also shown are three contour lines, corresponding to characteristic values of the unperturbed
mass density $\rho_0$ given by equation (\ref{eq:rho0}). The first contour line is for
$\rho_1 = 1.15\cdot\rho_0(\tilde{z}=0)$,
i.e. $15$ percent larger than the maximal value of $\rho_0$. The second contour line is for
the mass density value $\rho_2 = \rho_0(\tilde{z} = \tilde{c}_{z0})$. For low optical depths
$\tilde{c}_z$ is a good estimate of $\tilde{z}_\text{rms}$, the root mean square value of the
vertical excursion of the ring particles. The third contour line shows the mass density
$\rho_3 = \rho_0(\tilde{z} = \sqrt{3}\tilde{c}_{z0})$, where $z = \sqrt{3}\tilde{c}_z$
is half of the effective geometric thickness of the ring. The effective geometric thickness
$H_\text{eff} = \sqrt{12}\tilde{z}_\text{rms} \approx \sqrt{12}\tilde{c}_z$ of the ring corresponds
to the width of a uniform vertical mass density profile with the same standard deviation
as a Gaussian one.

Figure \ref{fig:zsq} shows different measures of the ring thickness after the
scattering by the moonlet. To calculate the ring thickness from the mass density, we
consider the mass in bin $(j',k')$ divided by the total mass in bins with bin index
$j'$
\begin{equation}
q(j',k') = \frac{\rho(j', k')}{\sum_{k'} \rho(j',k')}\ ,
\end{equation}
where we assumed uniform bin sizes. As a quantity, $q$ describes the distribution of
mass in the vertical direction and formally behaves like a probability, i.e.
$0 \le q(j',k') \le 1$ and $\sum_{k'} q(j',k') = 1$. The vertical displacement of the
ring plane from $\tilde{z} = 0$, calculated from the mass density, is then given by
\begin{equation}
\tilde{\mu}_{z,j'} = \sum_{k'} \bar{z}_{k'}\,q(j', k')\ ,
\end{equation}
with $\bar{z}_{k'}$ being the midpoint of the vertical interval $(\tilde{z}_{k'}, \tilde{z}_{k'+1})$.
The scale height of the ring is then proportional to
\begin{equation}
\tilde{\sigma}_z(j') = \left( \sum_{k'} \left(\bar{z}_{k'} - \tilde{\mu}_{z,j'} \right)^2 q(j', k') \right)^{1/2}\ .
\end{equation}

We also calculated the root mean square value of the vertical excursion of ring
particles directly from the results of test particle integrations. With the
averaging procedure of equation (\ref{eq:orbit_averaging}) $\tilde{z}_\text{rms}$
is given by
\begin{equation}
\tilde{z}_{\text{rms},j'} = \left( \frac{\sum_n p^{(n)}(j')\,[\tilde{z}^{(n)} - \langle\tilde{z}\rangle_{j'}]^2}
{\sum_n p^{(n)}(j')} \right)^{1/2}\ ,
\end{equation}
with
\begin{equation}
\langle\tilde{z}\rangle_{j'} = \frac{\sum_n \tilde{z}^{(n)}\,p^{(n)}(j')}
{\sum_n p^{(n)}(j')}\ .
\end{equation}

For the unperturbed mass density $\rho_0$, both $\tilde{\sigma}_{z0}$ and
$\tilde{z}_{\text{rms}0}$ equal $\tilde{c}_{z0}$. The values after the scattering
by the moonlet shown in Figure \ref{fig:zsq} are within $6$ percent of the values
of the unperturbed ring. Calculations for moonlets with different Hill radii from
$50$m to $500$m confirm that the difference of the unperturbed values to the ones
after the scattering by the moonlet are small, e.g. a few percent.

Up to now we have only considered the gravitational interaction with the moonlet.
Thus, the values of $\tilde\sigma_z$, determined from the mass density $\rho$,
and of $\tilde{z}_\text{rms}$, the root mean square of the test particle's vertical
excursions, are the result of gravitational excitement only. The increase of a few
percent of these values can not explain the height of propeller structures as infered
from their shadow length. A picture which changes drastically if a moonlet on an
inclined orbit is considered, which is an issue of ongoing future work.

Because the moonlet does not induce considerable vertical excursions of ring particles,
we restrict, in the following, our description to the ring plane and model the vertical
propeller structure by the granular ring temperature
$T = (\tilde{c}_x^2 + \tilde{c}_y^2 + \tilde{c}_z^2) / 3$.

\subsubsection{Velocity dispersion}
\label{sec:thermal_velocities}

\begin{figure}
\begin{adjustwidth}{-1in}{-1in}%
\begin{subfigure}[t]{0.7\textwidth}
    \centering
    \includegraphics[width=\textwidth]{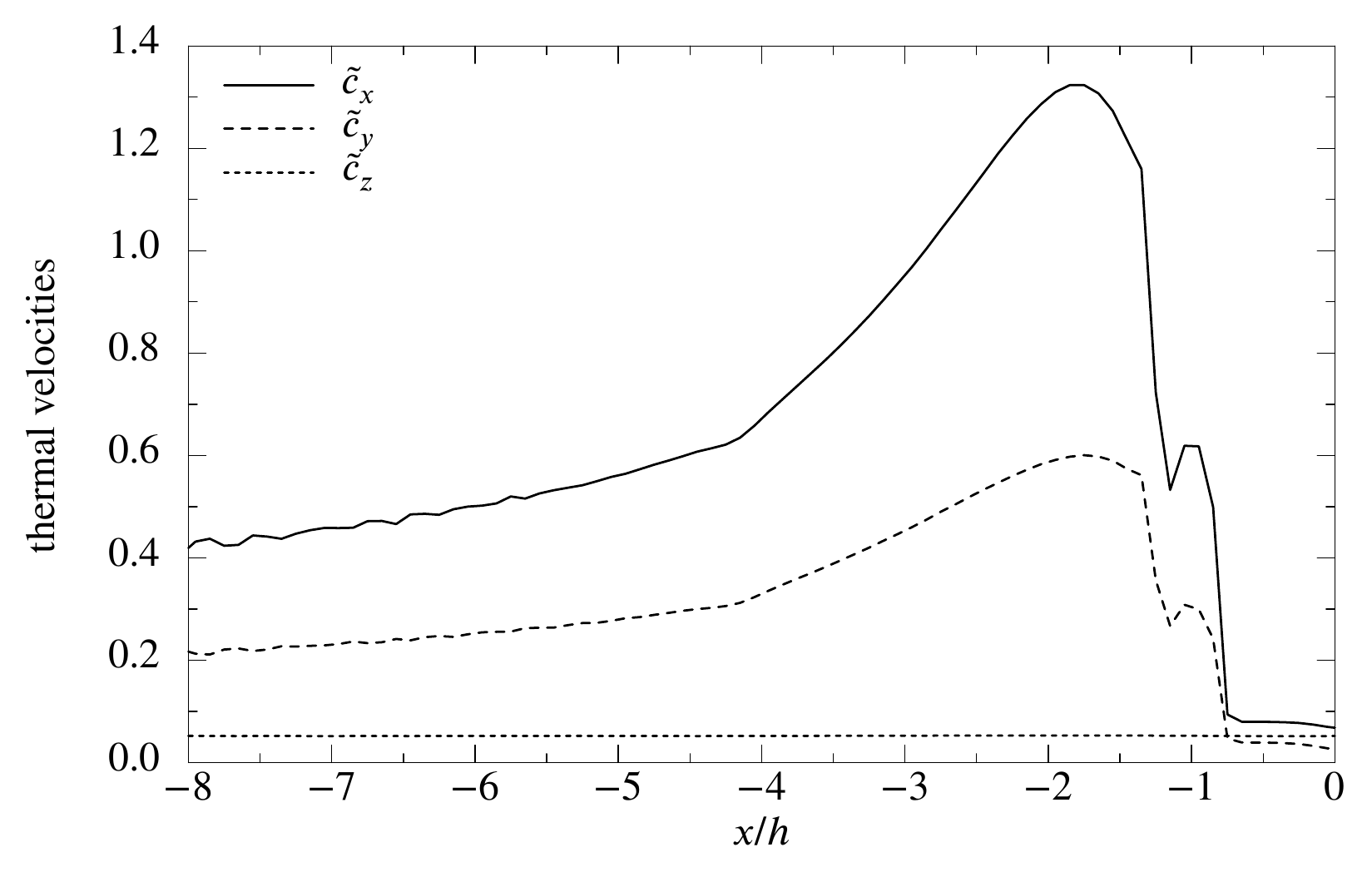}
    \caption{}\label{fig:cxyz}
\end{subfigure}%
\qquad
\begin{subfigure}[t]{0.7\textwidth}
\includegraphics[width=\textwidth]{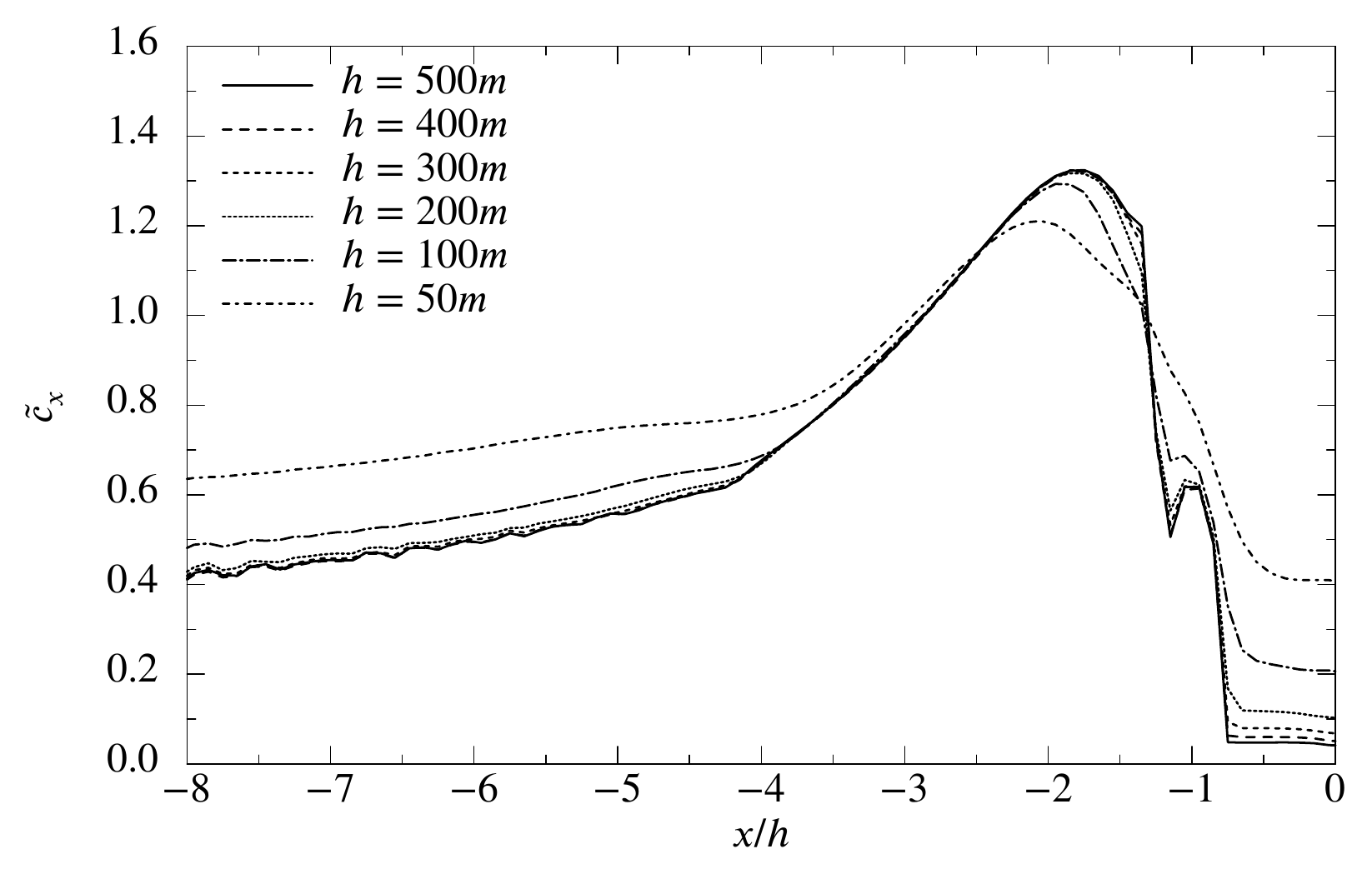}
\caption{}\label{fig:cx_scale}
\end{subfigure}
\caption{Moonlet induced thermal velocities as functions of the radial coordinate $\tilde{x}$ at
         $\tilde{y}=0^+$. All thermal velocities are scaled by $\Omega_0 h$. (a) Thermal velocities
         after the gravitational scattering by a moonlet with $300$m Hill radius. The moonlet induces
         much more thermal motion in the ring plane than in vertical direction. (b) Comparison of the
         radial thermal velocity $\tilde{c}_x$ after the scattering by moonlets with different Hill
         radii. In the gap region $-4 \le \tilde{x} \le -1.5$ the radial thermal velocity scales well
         with $\Omega_0 h$ for the larger moonlets ($h=100$m and above).}\label{fig:velodisp}
\end{adjustwidth}
\end{figure}


In this subsection we determine values of the moonlet induced thermal velocities after
the scattering by the moonlet, which are later used as initial conditions for the
hydrodynamical equations. We assume that after the scattering by the moonlet, the particles
are on Keplerian orbits which are well described by equation (\ref{eq:homhill}). We
calculate the thermal speed components $\tilde{c}_x$, $\tilde{c}_y$ and $\tilde{c}_z$
by taking the weighted sample standard deviation of $\tilde{v}_x$, $\tilde{v}_y - 3\tilde{x}/2$
and $\tilde{v}_z$ for the set of integrated test particles. For $\tilde{c}_x$ for example
\begin{equation}
\tilde{c}_x(j') = \left( \frac{\sum_n p^{(n)}(j')\,[\tilde{v}_x^{(n)} - \langle\tilde{v}_x\rangle(j')]^2}
{\sum_n p^{(n)}(j')}\right)^{1/2}\ ,
\label{eq:velodisp}
\end{equation}
where $j'$ denotes the radial bin number.

Due to the averaging process (\ref{eq:orbit_averaging}) our model does not describe
moonlet wakes, which are moonlet induced coherent motions of ring particles
\citep{showalter86,spahn94}. Nevertheless our model describes the gap region quite well,
because test particles ending in this region are often on trajectories which are very sensitive to
initial conditions \citep{henon86}, and their phases are mixed up. Furthermore, wake
structures start to dissolve when nearby streamlines cross.  For the gap region this
happens very fast, i.e. for the middle of the gap at $|\tilde{x}| = 2.5$ after $0.4$ orbits.

%
%
%

For the region beyond the gap however, we have to keep in mind that we might overestimate
$\tilde{c}_x$ and $\tilde{c}_y$ with equation (\ref{eq:velodisp}). The averaging process destroys
the coherent phase relations and systematic motion looks partly like thermal motion. In
this work we focus on the moonlet induced gap, which is important on its own, because of
the large thermal excitation in the gap region (cf. Figure \ref{fig:velodisp}).

The values of the mean thermal velocities, calculated with equation (\ref{eq:velodisp}), are the
result of the gravitational interaction with the moonlet only. Figure \ref{fig:zsq} shows
$\tilde{c}_z$ and $\tilde{z}_\text{rms}$, which do not deviate much at $\tilde{y}=0^+$. In Figure
\ref{fig:cxyz} we compare the mean thermal velocities after the scattering by the moonlet
of Figure \ref{fig:massdensities}.  The maximum excitement in each case lies near the inner
edge of the moonlet induced gap at a radial position of about $|\tilde{x}|=7/4$.


The moonlet induces much more thermal motion in the ring plane than in vertical direction.
This can be seen in the ratio $c_z/c_x$ after the scattering, which is several times
smaller than the equilibrium value of $0.65$. The ratio of the horizontal components
$c_y/c_x$, however, is close to the equilibrium value of $0.5$. Therefore, although the
graviational interaction with the moonlet induces thermal motion in vertical direction, it
is negligible compared to the thermal motion induced in the ring plane for moonlets on
non-inclined orbits.


Figure \ref{fig:cx_scale} shows the x-component of the velocity dispersion, scaled by $\Omega_0 h$,
at $\tilde{y} = 0^+$ after the scattering by moonlets with Hill radii of $50$m to $500$m.
For the larger moonlets, with a Hill radius above $100$m, the x-component of the velocity
dispersion in the gap region, especially the maximal value of $\tilde c_x$, scales well with the Hill
radius of the moonlets.

\subsection{Hydrodynamic Flow}
\label{sec:hydrodynamic_flow}

We describe the ring outside of the scattering region with hydrodynamical equations. Because
of the vertically ineffective gravitational scattering of ring particles by the moonlet, we
describe the thickness of the ring through the ring temperature, and use vertically averaged balance
equations. The mass and momentum balance are given by
\begin{align}
\frac{\partial \Sigma}{\partial t} + \nabla\cdot (\Sigma\mathbf{u}) &= 0 \\
\frac{\partial\mathbf{u}}{\partial t} + (\mathbf{u}\cdot\nabla) \mathbf{u}
  &= \mathbf{f} - \frac{1}{\Sigma}\nabla\circ\mathbf{P}\ .
\end{align}
Here $\Sigma$, $\mathbf{u}$, $\mathbf{f}$ and $\mathbf{P}$ are the surface mass density, the
mean velocity, external volume forces, and the pressure tensor. The pressure tensor is given
in Newtonian form
\begin{equation}
\mathbf{P} = p\mathbf{I} - 2\Sigma\nu\mathbf{D} - \Sigma\zeta (\nabla\cdot\mathbf{u})\mathbf{I}\ ,
\end{equation}
where $\mathbf{I}$ is the unit tensor. The vertically integrated pressure, kinematic bulk and
shear viscosity are denoted by $p$, $\zeta$ and $\nu$. The traceless shear tensor $\mathbf{D}$
is given by
\begin{equation}
\mathbf{D} = \frac{1}{2} \left(
   \nabla\circ\mathbf{u} + \mathbf{u}\circ\nabla
      \right) - \frac{1}{3} (\nabla\cdot\mathbf{u}) \mathbf{I}\ ,
\end{equation}
where $(\nabla\circ\mathbf{u})_{jk} = \nabla_j u_k$ and $(\mathbf{u}\circ\nabla)_{jk} = \nabla_k u_j$.

The energy balance of the ring particle's random motion reads
\begin{equation}
\frac{3}{2} \Sigma \left( \frac{\partial T}{\partial t} + (\mathbf{u}\cdot\nabla) T \right)
  = - \mathbf{P}:\mathbf{\epsilon} - \nabla\cdot\mathbf{Q} - \Gamma\ ,
\end{equation}
where $T = (c_x^2 + c_y^2 + c_z^2) / 3$ is the granular ring temperature and $\mathbf{Q} = -\kappa_D\nabla T$
the heat flow in the ring with heat conductivity $\kappa_D$. The friction term $\mathbf{P}:\mathbf{\epsilon}$
describes the viscous heating of the ring, and $\Gamma$ accounts for the energy loss due to inelastic collisions.

\subsubsection{Mass and momentum balance}
\label{sec:mass_balance}

After the gravitational scattering by the moonlet has opened a gap, the diffusion of the ring particles,
described by the nonlinear viscous diffusion equation
\begin{equation}
\frac{\partial \Sigma}{\partial t} +
\left( \Omega(r) - \Omega_0 \right) \frac{\partial \Sigma}{\partial\varphi} - 
\frac{3}{r} \frac{\partial}{\partial r}
\left[ \sqrt{r} \frac{\partial}{\partial r} \left( \sqrt{r}\nu\Sigma \right) \right] = 0\
\label{eq:diffusion}
\end{equation}
for the surface mass density $\Sigma$, will smooth out the induced structures \citep{spahn00, sremcevic02}.

In order to solve equation (\ref{eq:diffusion}) we apply the following simplifications, already used by
\citet{sremcevic02}:
Let $\sigma_1 = \Sigma - \Sigma_0$, where $\Sigma_0$ is the equilibrium value of
the surface mass density of the unperturbed ring. In the special case of constant kinematic
viscosity $\nu_0$ and without curvature terms, which is consistent with the Hill approximation
of the gravitational scattering, equation (\ref{eq:diffusion}) reduces to a linear partial
differential equation
\begin{equation}
K \tilde{x} \frac{\partial\sigma_1}{\partial\varphi} = 
   - \frac{\partial^2 \sigma_1}{\partial \tilde{x}^2}\ ,
\label{eq:lin_diffusion}
\end{equation}
where the constant $K$ is defined by
\begin{equation}
K = \frac{\Omega_0 a_0^2}{2\nu_0} \left( \frac{h}{a_0} \right)^3\ .
\end{equation}
%


Equation (\ref{eq:lin_diffusion}) is point symmetric with respect to the position of
the moonlet ($\tilde{x} = \tilde{y} = \tilde{z} = 0$). For the region defined by
$\tilde{x} < 0$ and $\varphi > 0$, with boundary conditions
\begin{align}
\sigma_1(\tilde{x} < 0, \varphi = 0^+) &= \sigma_{1,\text{ini}} \\
\sigma_1(\tilde{x} \to -\infty, \varphi) &= 0 \\
\sigma_1(\tilde{x}, \varphi \to \infty) &= 0 \ ,
\end{align}
\citet{sremcevic02} derived several Greens functions for different constraints
at $\tilde{x} = 0$, $\varphi > 0$. The Greens function, which matched their numerical
solution best, is an equally weighted superposition of Greens functions for the two
cases
\begin{align}
\sigma_1(\tilde{x} = 0, \varphi) &= 0 \\
\frac{\partial\sigma_1}{\partial\tilde{x}} (\tilde{x} = 0, \varphi) &= 0\ .
\end{align}

\begin{figure}
\begin{adjustwidth}{-1in}{-1in}%
\begin{subfigure}[t]{0.7\textwidth}
    \centering
    \includegraphics[width=\textwidth]{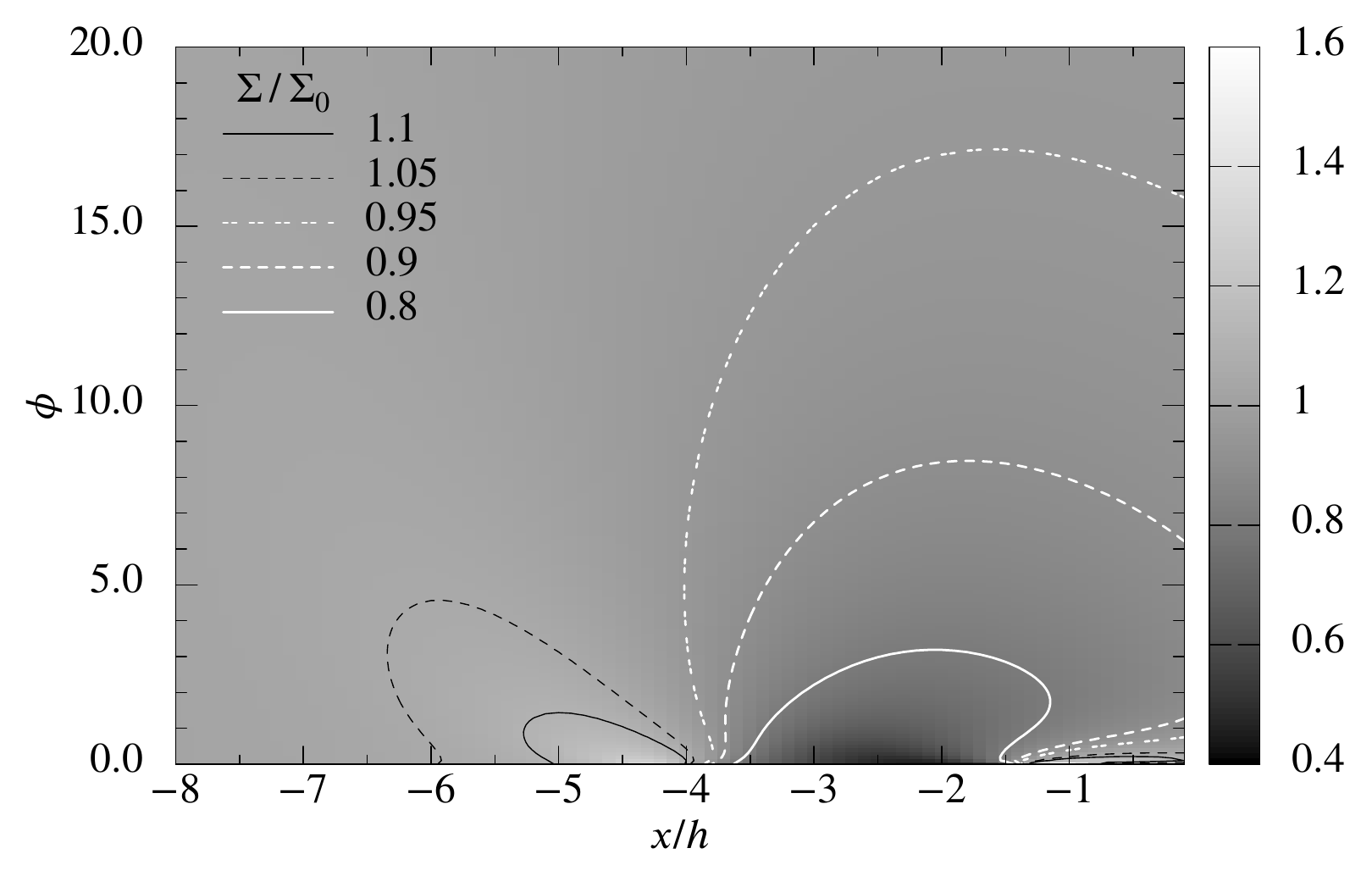}
    \caption{}\label{fig:density_cnt}
\end{subfigure}%
\qquad
\begin{subfigure}[t]{0.7\textwidth}
    \centering
    \includegraphics[width=\textwidth]{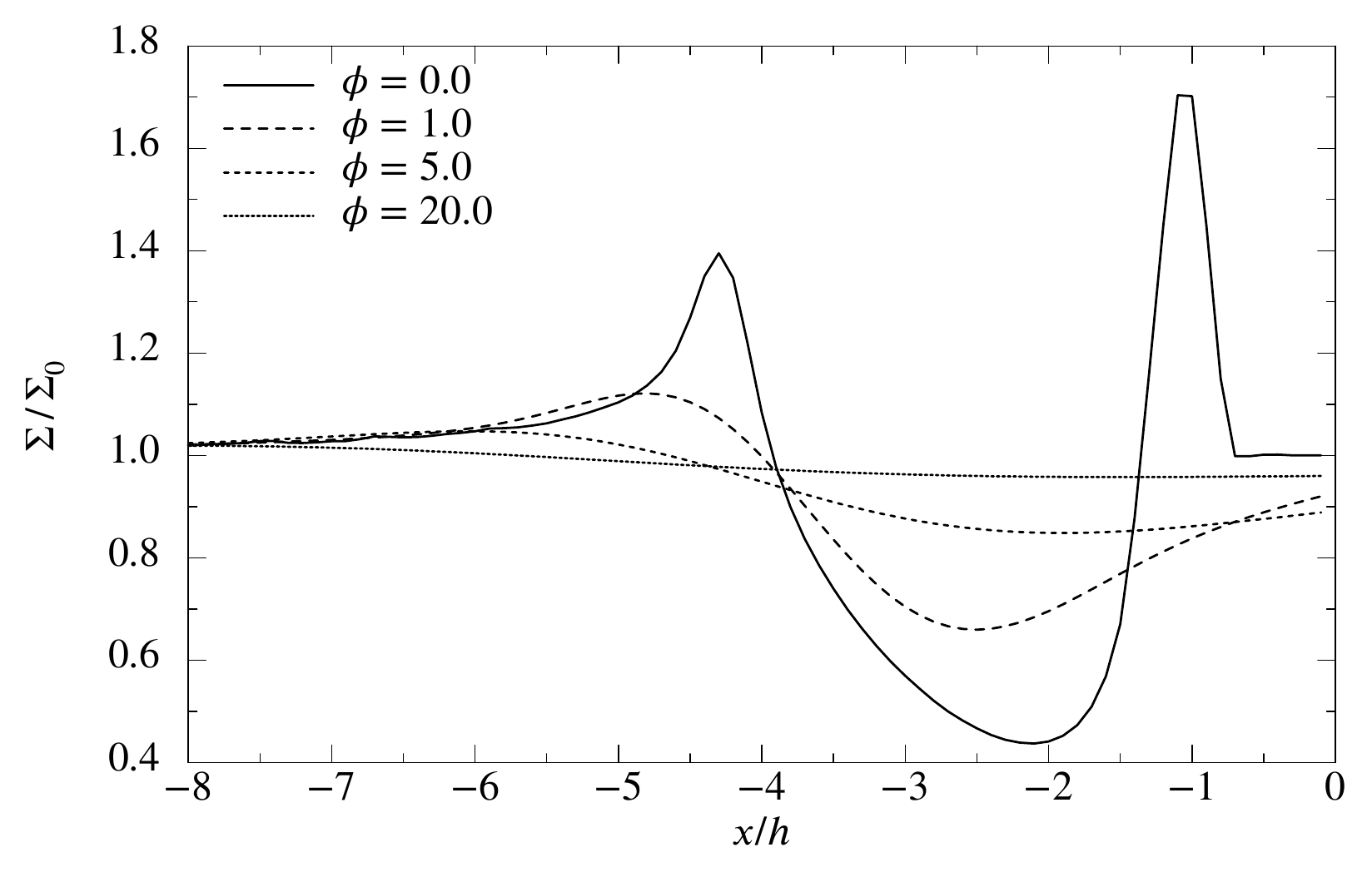}
    \caption{}\label{fig:density_phi}
\end{subfigure}%
\caption{(a) Contour plot of the surface mass density downstream of the moonlet. (b) Surface
         mass density as a function of $\tilde{x}$ for different longitudes, illustration the
         gap-closing. The solid line shows the surface mass density after the scattering by the
         moonlet.}
         \label{fig:massdensity}
\end{adjustwidth}%
\end{figure}

We use this Greens function, given by
\begin{equation}
G(\tilde{x}, \phi) = \frac{\sqrt{3}}{2} (-\tilde{x}_0) \left(3\phi\right)^{2/3}
\ \exp\left( \frac{\tilde{x}_0^3 + \tilde{x}^3}{9\phi} \right)
\text{Bi} \left( (3\phi)^{2/3} \tilde{x}_0 \tilde{x} \right)\ ,
\end{equation}
with the scaled azimuth $\phi = \varphi / K$, and the Airy function $\text{Bi}(z)$,
to calculate the surface mass density via
\begin{equation}
\sigma_1(\tilde{x}, \phi) = \int_{-\infty}^0 d\tilde{x}_0\, \sigma_{1,\text{ini}}(\tilde{x}_0)\, G(\tilde{x}, \phi; \tilde{x}_0)\ .
\label{eq:mass_relax}
\end{equation}
As initial surface mass density $\sigma_{1,\text{ini}}$ we use the vertically integrated mass
density calculated with the scattering operator (\ref{eq:dens_calc}), shown in Figure
\ref{fig:density_phi} as solid line, and numerically integrate equation (\ref{eq:mass_relax}). The two
plots in Figure \ref{fig:massdensity} are again made for a moonlet with $300$m Hill radius and
the orbital parameters of Earhart. For the value of the kinematic viscosity in Saturn's A ring
we used $\nu_0 = 0.01$ $\text{m}^2/$s \citep{tiscareno07}.

Figure \ref{fig:density_cnt} shows the surface mass density after the scattering by the moonlet
as function of $\tilde{x}$ and $\phi$. White contour lines represent surface mass density values
below $\Sigma_0$, showing the evolution of the gap. Black contour lines enclose regions of enhanced
surface mass density compared to $\Sigma_0$. To illustrate the gap-closing, we plotted the surface
mass density as a function of the radial coordinate $\tilde{x}$ for different longitudes $\phi$,
shown in Figure \ref{fig:density_phi}.

%
The azimuthal extent of the surface mass density scales with $a_0 K$. Because $K\propto h^3$,
the importance of the mass diffusion process, in the first few orbits after the scattering by
the moonlet, depends strongly on the Hill radius of the moonlet. For large moonlets, e.g.
$h=500$m, the surface mass density stays close to the initial value at $\phi=0$. The smaller
the moonlet the larger the influence of the mass diffusion in the first few orbits.

\subsubsection{Energy balance}
\label{sec:energy_balance}

In this subsection we will consider the relaxation of the moonlet induced thermal heating
by dissipative particle interaction. The vertically ineffective gravitational interaction
of ring particles with the moonlet can not explain the vertical height of propeller structures.
The moonlet induces thermal motion mainly in lateral direction.

But due to collisions between ring particles, the thermal motion induced in lateral direction
will be converted to vertical thermal motion till the ratio $c_z / c_x$ reaches its local
equilibrium value. This asymptotic value of $c_z / c_x$ is established very fast, after a few
collisions per particle \citep{haemeen80, haemeen93}. We therefore use the granular temperature
$T$ to model the evolution of the vertical propeller structure, infering the thickness of
the ring from $T$ and the equilibrium value of $c_z / c_x$.


%

The thermal energy balance equation in the corotating frame reads
\begin{equation}
\frac{3}{2} \Sigma \left\{
\frac{\partial T}{\partial t}
+ (\Omega - \Omega_0) \frac{\partial T}{\partial\varphi}
+ u_r \frac{\partial T}{\partial r} \right\} = 
- \mathbf{P}:\mathbf{\epsilon} - \nabla\cdot\mathbf{Q} - \Gamma\quad.
\end{equation}
%

The temperature relaxation of the ring is much slower than the relaxation of the ratio
$c_z/c_z$ to its local equilibrium value. On the other hand, the temperature relaxation
is fast compared to the mass diffusion timescale. Therefore we consider, for simplicity,
only viscous heating and energy loss due to inelastic collisions.  Furthermore, we assume
stationarity to obtain
\begin{equation}
\frac{3}{2} (\Omega - \Omega_0) \frac{\partial T}{\partial\varphi} = 
\frac{9}{4}\,\nu \Omega_0^2 - k_3 \Omega_0 \tau (1-\varepsilon^2) T\ ,
\label{eq:temperatur_simplified}
\end{equation}
where we used $\mathbf{P}:\mathbf{\epsilon} = - 9\nu \Omega_0^2\Sigma/4$ for the
friction term, regarding just the dominant Kepler shear. For the cooling term we
used $\Gamma = k_3 \Omega_0 \tau (1-\varepsilon^2) \Sigma T$ with a constant coefficient
of restitution $\varepsilon = 0.5$ and with $k_3 = 1.5$ \citep{stewart84}, corresponding
to an energy loss rate
\begin{equation}
\dot{E}_\text{coll} = -\omega_c/6\,(1-\varepsilon^2)\,\Sigma c^2\ ,
\end{equation}
with collison frequency $\omega_c = 3 \Omega\tau$, and temperature $c^2 = 3T$.

We assume the viscosity to be constant $\nu=\nu_0$ on timescales which are relevant
for the exponential decay of the temperature. With a constant viscosity and a constant
coefficient of restitution, viscous heating and collisional cooling balance at a local
equilibrium temperature
%
\begin{equation}
T_\text{eq}(\tau) = \frac{9 \nu_0 \Omega_0}{4 k_3 (1-\varepsilon^2) \tau}\ ,
\end{equation}
which depends on the local optical depth, or assuming equally sized ring particles with
particle radius $R_p$ and mass $m_p$, via $\tau = \pi R_p^2 \Sigma/m_p$, on the
surface mass density $\Sigma$. The temperature of the unperturbed ring before the
encounter with the moonlet is $T_0 = T_\text{eq}(\tau_0)$.

The temperature independent viscosity seems to be a strong restriction and is
considerd to be a first step towards a more complete description of the problem.
The applicability of our model presented in Section~\ref{sec:discussion} will
justify the simplifications applied in this work.

With these assumptions equation (\ref{eq:temperatur_simplified}) becomes a linear
ordinary differential equation and with
$(\Omega(r) - \Omega_0)\,\partial/\partial\varphi = -3/2 \Omega_0 \tilde{x}\,\partial/\partial\tilde{y}$
and $T_1 = T - T_0$:
\begin{equation}
\tilde{x}\ \frac{\partial T_1}{\partial\tilde{y}} = 
\nu_0 \Omega_0 \frac{\sigma_1}{\Sigma_0} + \frac{4}{9} k_3 \tau (1-\varepsilon^2) T_1\ .
\label{eq:temperatur_further_simplified}
\end{equation}
The solution of this equation can be written in the form
\begin{equation}
T_1(\tilde{x}, \tilde{y}) = T_{1,\text{ini}}(\tilde{x})\, e^{-\gamma(\tilde{x}, \tilde{y})}
  + e^{- \gamma(\tilde{x}, \tilde{y})}
    \int_0^{\tilde{y}} f(\tilde{y}'') e^{\gamma(\tilde{x}, \tilde{y}'')}\,d\tilde{y}'',
\label{eq:tsolution}
\end{equation}
with the functions
\begin{align}
\gamma(\tilde{x},\tilde{y}) &= \frac{4 k_3 (1-\varepsilon^2) \tau_0}{9 |\tilde{x}|}
     \int_0^{\tilde{y}} \frac{\Sigma(\tilde{x},\tilde{y}')}{\Sigma_0} d\tilde{y}' \label{eq:gamma_func} \\
f(\tilde{x}, \tilde{y}) &= - \frac{\nu_0 \Omega_0}{|\tilde{x}|}\,\frac{\sigma_1(\tilde{x},\tilde{y})}{\Sigma_0}\ .\nonumber
\end{align}

\begin{figure}
\begin{adjustwidth}{-1in}{-1in}%
\begin{subfigure}[t]{0.7\textwidth}
    \centering
    \includegraphics[width=\textwidth]{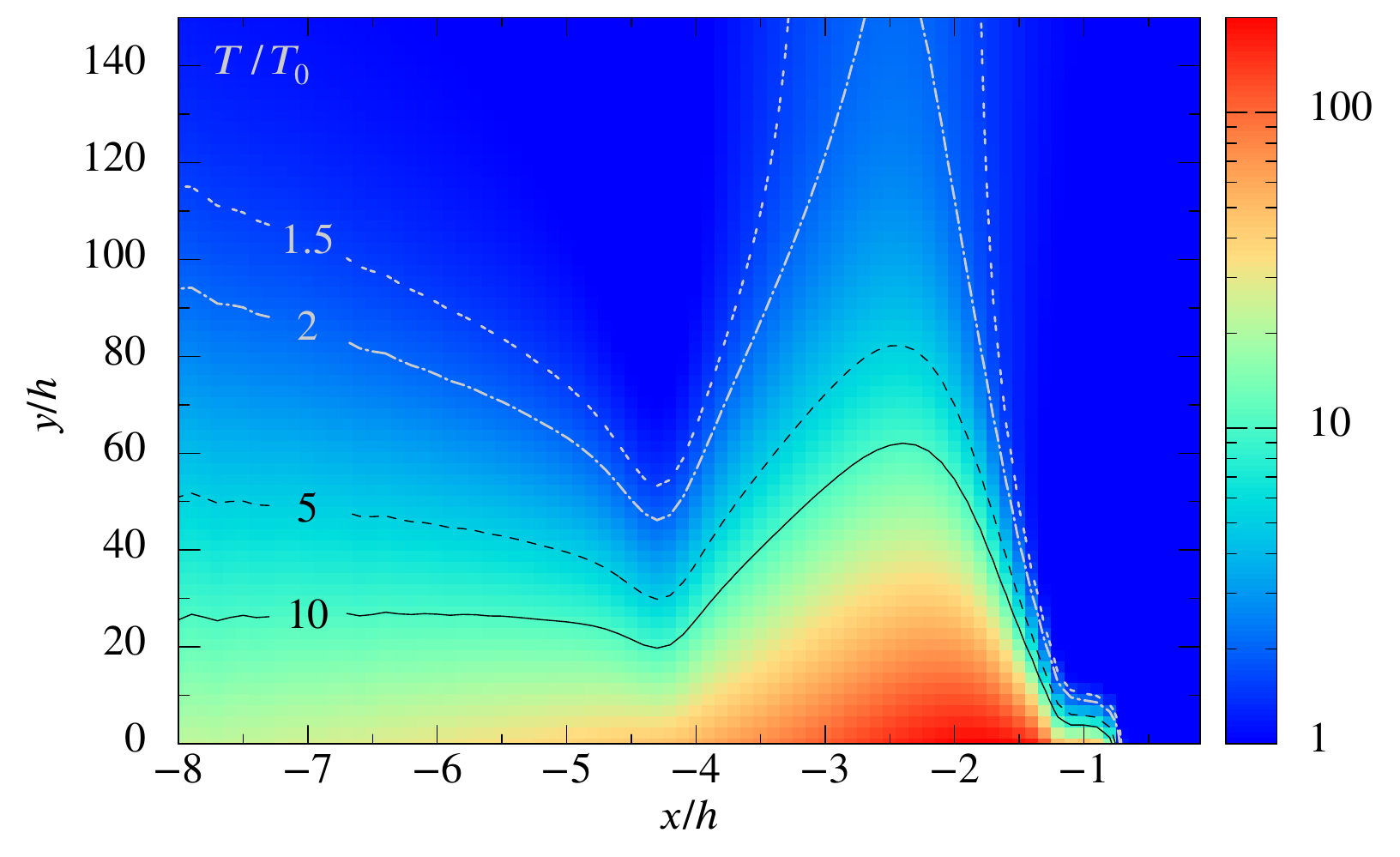}
    \caption{}\label{fig:temp_cnt}
\end{subfigure}%
\qquad
\begin{subfigure}[t]{0.7\textwidth}
    \centering
    \includegraphics[width=\textwidth]{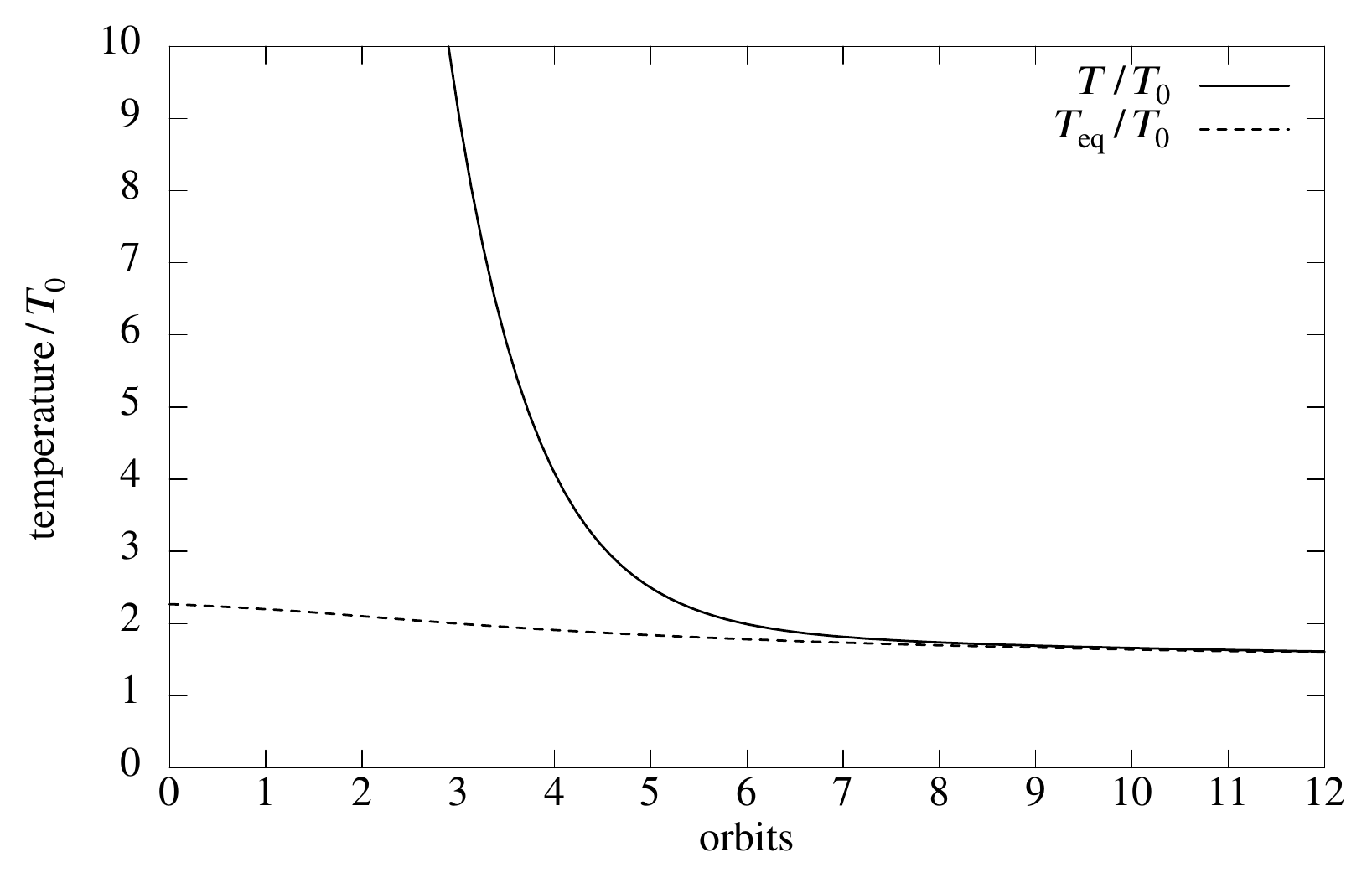}
    \caption{}\label{fig:temprelaxation}
\end{subfigure}%
\caption{Downstream temperature relaxation for a moonlet with a $300$m Hill radius. (a)
         Contour plot of the temperature decrease downstream of the moonlet. (b) Temperature
         relaxation at radial position $\tilde{x}=-2$. The solid line shows the temperature
         relaxation of the ring, the dashed line corresponds to the equilibrium temperature
         $T_\text{eq}$, for which viscous heating and collisional cooling are in balance.}
         \label{fig:temp_relax}
\end{adjustwidth}%
\end{figure}

\begin{figure}
\begin{adjustwidth}{-1in}{-1in}%
\begin{subfigure}[t]{0.7\textwidth}
    \centering
    \includegraphics[width=\textwidth]{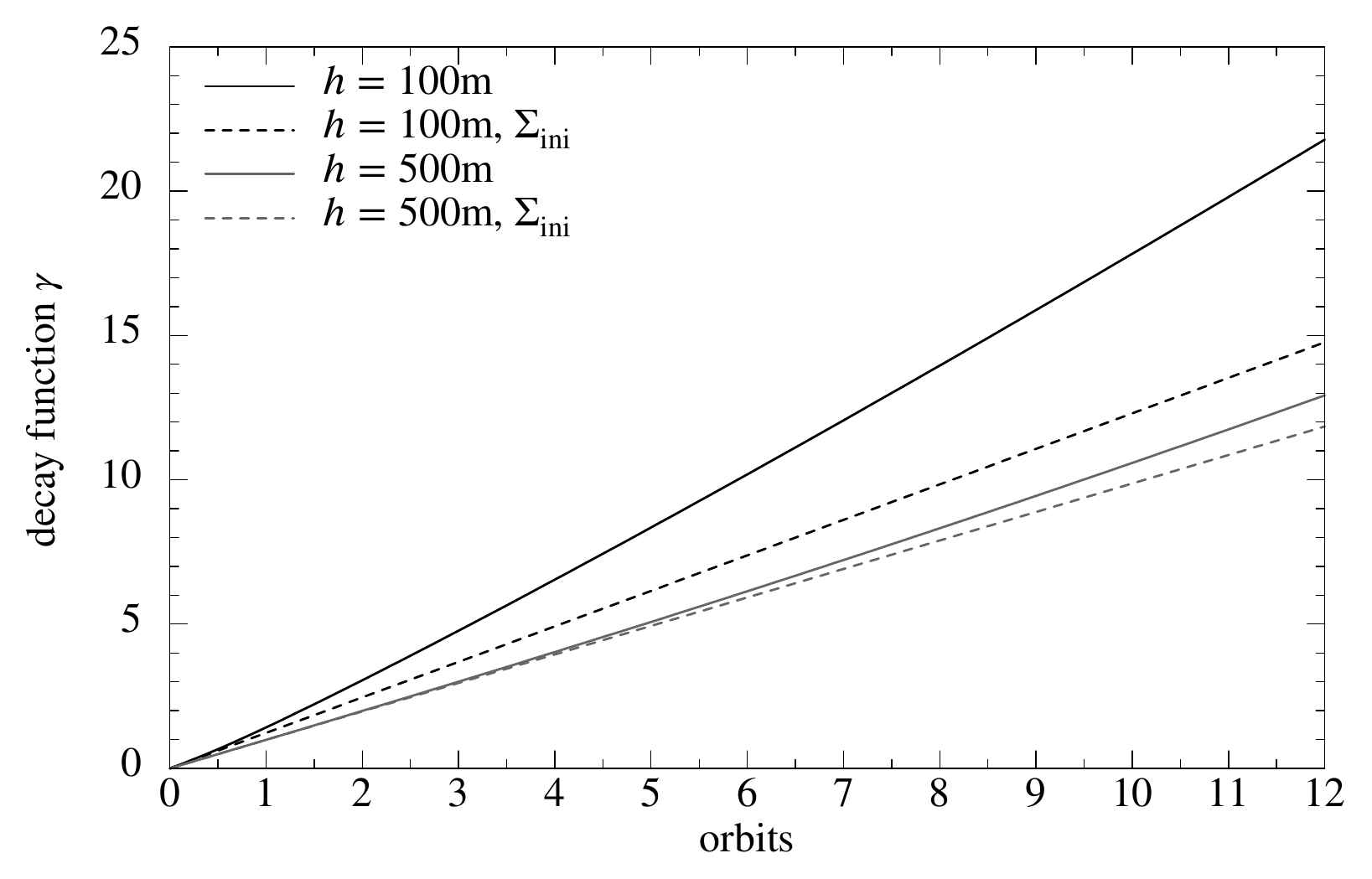}
    \caption{}
    \label{fig:gammacomp}
\end{subfigure}%
\qquad
\begin{subfigure}[t]{0.7\textwidth}
    \centering
    \includegraphics[width=\textwidth]{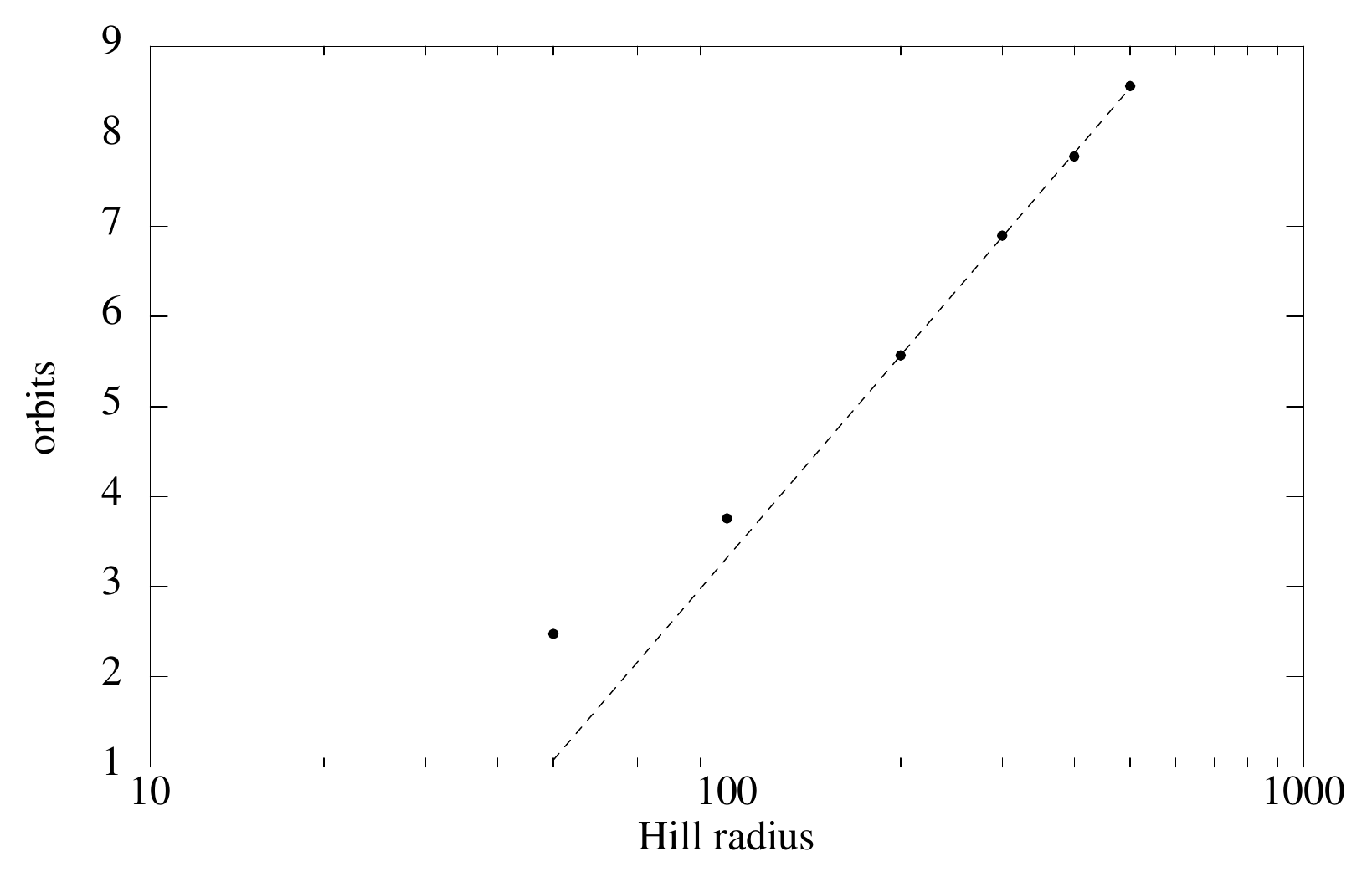}
    \caption{}
    \label{fig:ymax}
\end{subfigure}%
\caption{(a) Decay function $\gamma$ at fixed radial position $\tilde{x}=-2$ for two moonlets,
         one with $100$m Hill radius and one with $500$m Hill radius. The solid lines show
         $\gamma$ calculated according to equation (\ref{eq:gamma_func}). In contrast the dashed
         lines are calculated with an azimuthally constant surface mass density $\Sigma$ equal
         to $\Sigma_\text{ini}$. (b) It is shown how many orbits it takes the temperature to decay
         to $5$ percent of $T_\text{eq}$ for moonlets with different Hill radii at $\tilde{x} = -2$.}
         \label{fig:gamma_scaling}
\end{adjustwidth}%
\end{figure}


The plots in Figure \ref{fig:temp_relax} show the downstream temperature relaxation. Again
they are made for a moonlet with $300$m Hill radius and the orbital parameters of Earhart,
using $\nu_0 = 0.01\,\text{m}^2/\text{s}$ as value of the kinematic viscosity in Saturn's A
ring. Figure \ref{fig:temp_cnt} shows a contour plot of the ring temperature
$T$ scaled by the unperturbed ring temperature $T_0$, illustrating the azimuthal temperature
decay for different radial regions. Ring particles with $|\tilde{x}| < 0.75$ are on horseshoe
orbits and their minimal distance to the moonlet is large, so that there is no considerable
thermal excitation. The gap region, on the other hand, is highly excited for quite an azimuthally
extended range. Figure \ref{fig:temprelaxation} shows the azimuthal temperature evolution,
at fixed radial position $\tilde{x}=-2.0$, downstream from the moonlet. Also, the
local equilibrium temperature $T_\text{eq}$ is shown as function of the surface mass
density. For the first few orbits, the ring temperature decreases exponentially to the
local equilibrium temperature. Afterwards, the viscous heating and the collisional cooling
are balanced and the ring temperature evolves as a function of the surface mass density of
the ring. In this case the temperature gets within a margin of $5$ percent of $T_\text{eq}$
in about 7 orbits.

The influence of the mass diffusion on the temperature decay depends on the Hill radius of the
moonlet. Figure \ref{fig:gammacomp} shows decay functions $\gamma$, calculated at radial position
$\tilde{x}=-2$, for two moonlets with Hill radii of $100$m and $500$m. The solid lines show
$\gamma$ calculated according to equation (\ref{eq:gamma_func}). In contrast the dashed lines
are calculated with an azimuthally constant surface mass density $\Sigma$ equal to $\Sigma_\text{ini}$.
For the larger moonlet the solid and the dashed line stay fairly close to each other, because
the change of the surface mass density is rather small on the timescale of the exponential decay.
For the smaller moonlet both lines differ clearly indicating that the influence of mass diffusion
is by far larger.

A fair approximation to the solution (\ref{eq:tsolution}) is given by
\begin{equation}
T_\text{approx}(\tilde{x}, \tilde{y}) =
(T_\text{ini} - T_\text{eq}(\tau))\,e^{-\gamma(\tilde{x},\tilde{y})} + T_\text{eq}(\tau)\ .
\label{eq:tapprox}
\end{equation}
For $|\tilde{x}| \ge 1$ the maximal difference of equation (\ref{eq:tsolution}) to the above
approximate solution is about $5$ percent, tested for moonlets with Hill radii of $50$m
to $500$m.

The exponential decay of the temperature stops after a few orbits, when the local equilibrium temperature
$T_\text{eq}$ is reached. Figure \ref{fig:ymax} sketches how many orbits it takes to get within
$5$ percent of $T_\text{eq}$ for moonlets with different Hill radii at fixed radial position
$\tilde{x} = -2$. Using the approximate solution $T_\text{approx}$, the azimuthal position
$\tilde{y}_\xi$ where the the ratio $T/T_\text{eq}$ equals $1 + \xi$ is
\begin{equation}
\tilde{y}_\xi = \frac{1}{\langle \gamma \rangle}
\ln \left(
\frac{T_\text{ini} - T_\text{eq}(\tau)}{\xi\, T_\text{eq}(\tau)}
\right)\ ,
\label{eq:ymax}
\end{equation}
where we defined formally $\gamma(\tilde{x}, \tilde{y}) = \langle \gamma \rangle\,\tilde{y}_\xi$.
Both $\langle \gamma \rangle$ and $T_\text{eq}$ depend on the optical depth $\tau$ and therefore
on the azimuthal position $\tilde{y}$. For large moonlets, close in size to ring-moons able to open
a circumferential gap, $\tau$ is nearly constant on the timescale of the exponential temperature
decay. Because of $T_\text{ini} \gg T_\text{eq}$ and $T_\text{ini} \propto h^2$ for large moonlets,
$\tilde{y}_\xi$ will scale with $\tilde{y}_\xi \propto 2\ln(h)$. In Figure \ref{fig:ymax} a fit
of the function $\tilde{y}_\xi = A\,\ln(B h^2)$ to the values of the large moonlets with Hill radii
of $200$m to $500$m is presented in order to underline this scaling.

\section{The vertical height of propellers}
\label{sec:propeller_height}

\begin{figure}
\begin{adjustwidth}{-1in}{-1in}%
\begin{subfigure}[t]{0.7\textwidth}
    \centering
    \includegraphics[width=\textwidth]{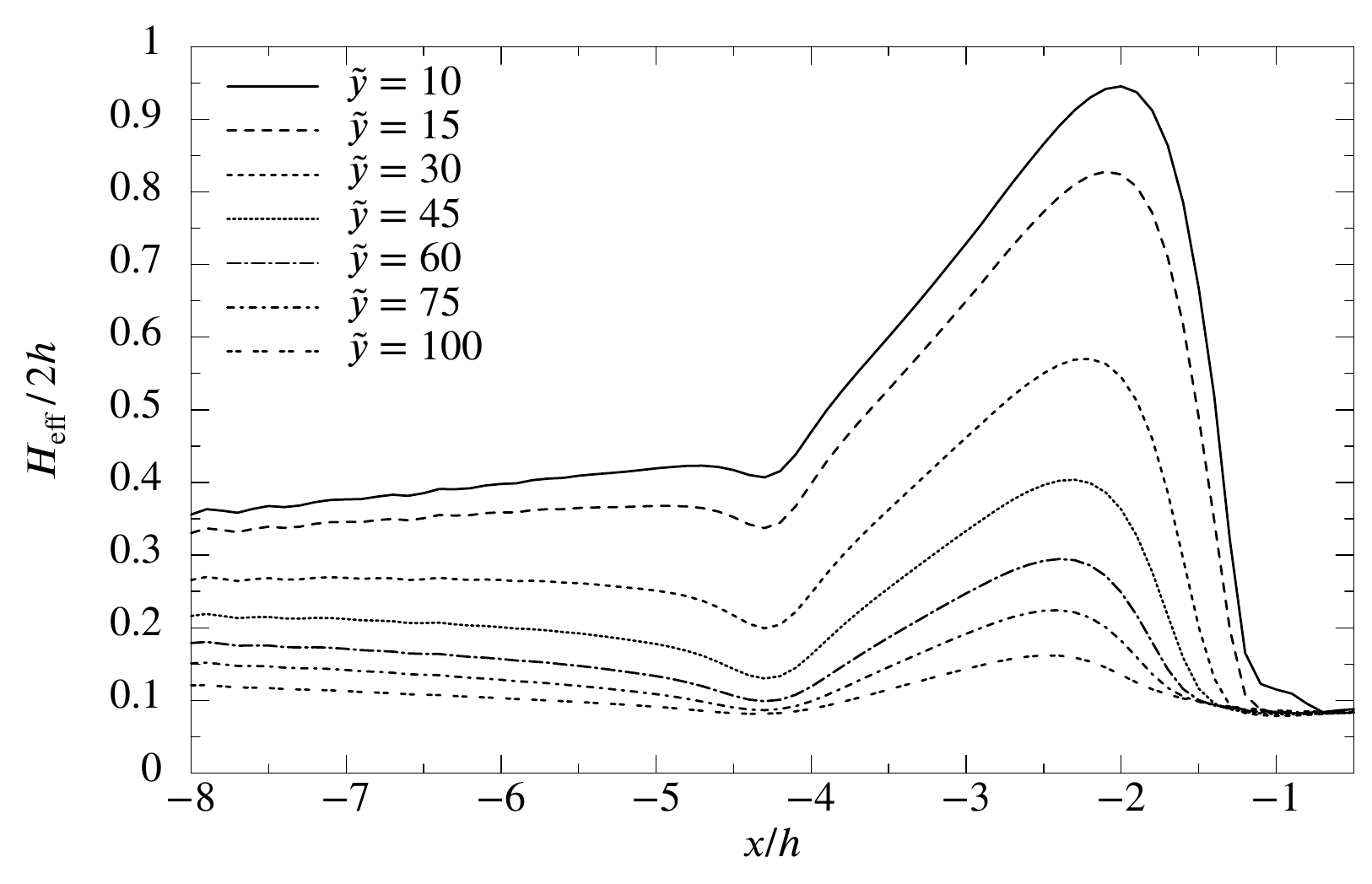}
    \caption{}\label{fig:height_ycuts}
\end{subfigure}%
\qquad
\begin{subfigure}[t]{0.7\textwidth}
    \centering
    \includegraphics[width=\textwidth]{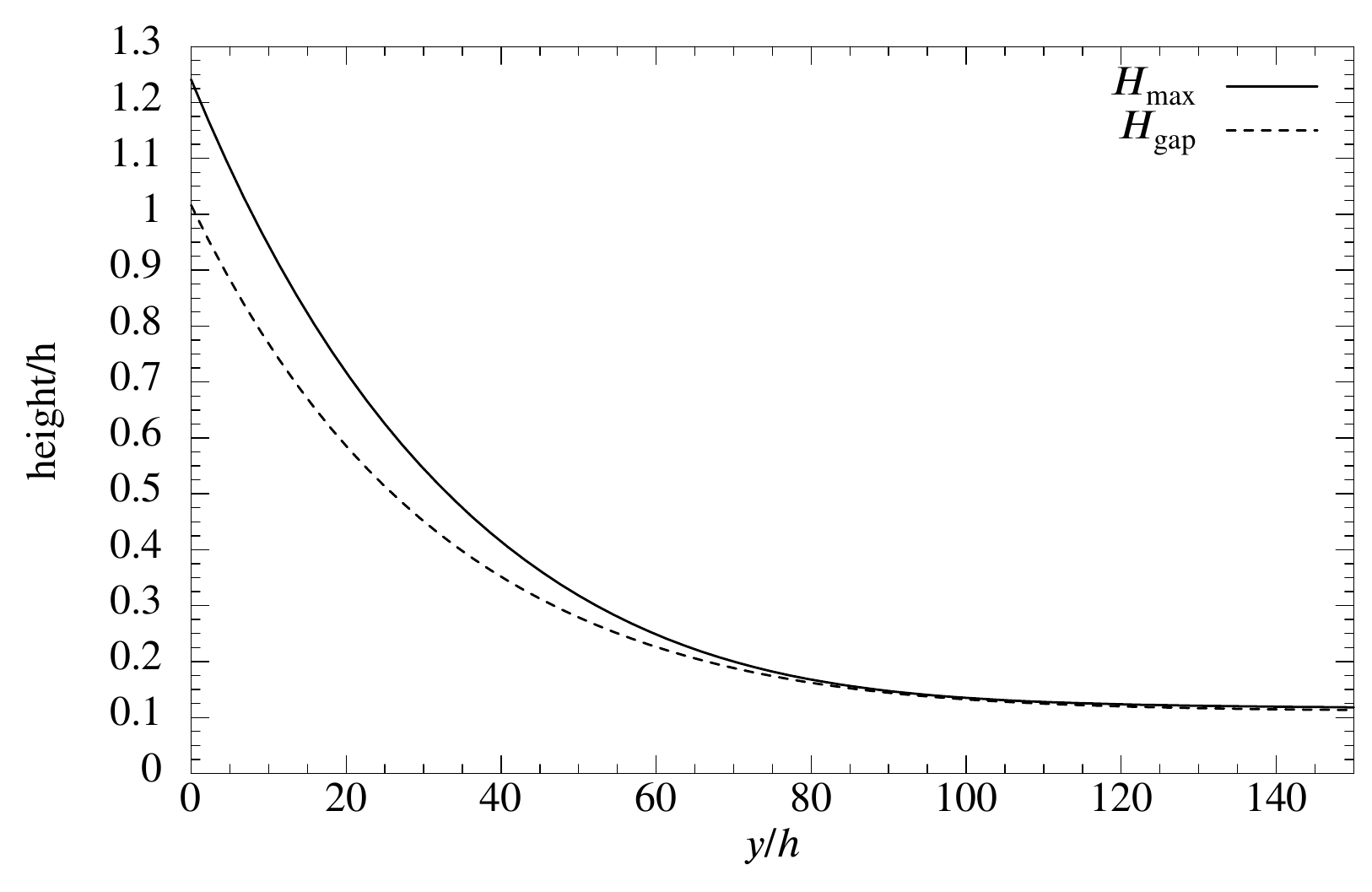}
    \caption{}\label{fig:max_height}
\end{subfigure}%
\caption{Azimuthal relaxation of the propeller height for a moonlet with $300$m Hill radius. (a)
         Propeller heights as function of radial coordinate $\tilde{x}$ for different longitudes.
         (b) Downstream relaxation of the propeller height. The solid line shows the relaxation
         at $\tilde{x} = -2$. The dashed line shows the propeller height radially averaged over the
         gap region.}\label{fig:height_av}
\end{adjustwidth}%
\end{figure}


In the thin-disk approximation and with a $z$ independent vertical velocity dispersion
$c_z$, the vertical profile of the mass density can be described well by a Gaussian with
standard deviation $c_z/\Omega_0$ \citep{stewart84,simon94,schmidt99}.  We use the effective
geometric thickness $H_\text{eff}$
as a measure of the ring thickness and determine the height of the propeller structure as
half the effective geometric thickness of the ring.

To determine $c_z$ as a function of the temperature $T$, we use the equilibrium values of
the ratios $c_z/c_x$ and $c_y/c_x$. A perturbed ring will reach the equilibrium values of
these quantities after a few collisions per particle \citep{haemeen80, haemeen93}. In
Saturn's A ring with $\tau=0.5$ and with a collision frequency of $\omega_c = 3 \Omega\tau$,
there are about $10$ collisions per particle per orbit. We assume that after $5$ collisions,
corresponding to half an orbit, the equilibrium values of these ratios are established.
With the relaxed ratios $(c_z/c_x)_\text{eq} = 0.65$ and $(c_y/c_x)_\text{eq} = 0.5$, the z
component of the velocity dispersion in terms of $T$ is given by
\begin{equation}
c_z(T) = \left(\frac{c_z}{c_x}\right)_{\!\!\text{eq}} \sqrt{\frac{3T}{1 + (c_y/c_x)^2_\text{eq} + (c_z/c_x)^2_\text{eq}}}\ .
\end{equation}

The two plots in Figure \ref{fig:height_av} are made using the same parameters as were
taken for Figure \ref{fig:temp_relax}, i.e. a moonlet with $300$m Hill radius, the
orbital parameters of Earhart and a kinematic viscosity of $\nu_0 = 0.01$ $\text{m}^2/\text{s}$.
In Figure \ref{fig:height_ycuts}, the height of the propeller structure as a function of
the radial coordinate $\tilde{x}$ is plotted for different values of $\tilde{y}$. The
largest heights are reached in the region between $\tilde{x} = -4$ and $\tilde{x} =
-1$, the maximum being at about $\tilde{x} = -2$ for $\tilde{y}=10$ and slowly moving
to $\tilde{x}=-2.5$ at $\tilde{y}=100$.

The azimuthal relaxation of the height of the propeller structure, for $\tilde{x} = -2$,
is shown in Figure \ref{fig:max_height}. We assume that after half an orbit or $5$ collisions
per particle the equilibrium value of $c_z/c_x$ is established and that afterwards the ring
temperature describes the propeller height well. At radial position $\tilde{x} = -2$, half
an orbit corresponds approximately to $\tilde{y} = 10$. The height at $\tilde{y} = 10$ is
$0.95$ Hill radii or about $285$m.

Furthermore, Figure \ref{fig:max_height} sketches the propeller height radially averaged over
the gap region. We choose the approximate middle of the gap at $\tilde{x} = -2.5$ as a reference,
and in that case half an orbit orbit corresponds to $\tilde{y} = 12$. At that position the gap
averaged propeller height is $0.72$ Hill radii or about $215$m.

\begin{table}[t]
\centering
\begin{tabular}{rcccccc}
\toprule
$h/m$    &  $50$  & $100$  & $200$  & $300$  & $400$  & $500$ \\
\midrule
$H_\text{max}/h$ & $0.90$ & $0.90$ & $0.94$ & $0.95$ & $0.95$ & $0.96$ \\
$H_\text{gap}/h$ & $0.80$ & $0.72$ & $0.72$ & $0.73$ & $0.73$ & $0.74$ \\
\bottomrule
\end{tabular}
\caption{Heights for  moonlets with different Hill radii from $50$m to $500$m.
$H_\text{max}$ is the height at $\tilde{x} = -2.0$ and $\tilde{y} = 10$.
$H_\text{gap}$ is the height averaged over the gap region at azimuthal position
$\tilde{y} = 12$.}\label{table:heights}
\end{table}

Table \ref{table:heights} shows propeller heights for moonlets with different Hill
radii. $H_\text{max}$ is the height at radial position $\tilde{x} = -2$ and azimuthal
position $\tilde{y} = 10$. $H_\text{gap}$ is the height averaged over the gap region
at azimuthal position $\tilde{y} = 12$. The maximal heights are close to one Hill radius
for the tested moonlet sizes. The values of the gap averaged height for the moonlets
with Hill radius $100$m and above are approximatly $0.7$ Hill radii.
The averaged propeller height for the moonlet with a $50$m Hill radius is larger
($0.8$h), which is due to the large value of the unperturbed height of the ring scaled
by the Hill radius of the moonlet.

\section{Discussion}
\label{sec:discussion}

We study the vertical extent of propeller structures in Saturn's rings, focusing on the
propeller gap region. The effective geometric thickness $H_\text{eff}$ is used to describe
the propeller height as a function of the ring temperature, calculated using the equilibrium
ratios of the thermal velocities. A vertically constant ring temperature serves as a fair
assumption, because for the low optical depths in the propeller gap region the vertical
dependence of the ring temperature is rather weak \citep{schmidt99}.

In our model the azimuthal temperature decrease is caused dominantly by the disturbed
balance of viscous heating and collisional cooling. We assume a constant coefficient of
restitution for the cooling term and constant viscosity for the viscous heating term.
This is a simplification, but allows a semi-analytical solution of the azimuthal temperature
relaxation (some integrals have to be numerically evaluated), and most important, it is
consistent with images of propeller shadows taken by the Cassini spacecraft.

The restriction of a constant $\nu_0$ could be dropped by introducing e.g. a power law dependence
\citep{spahn00b},
\begin{equation}
\nu = \nu_0 \left(\frac{\Sigma}{\Sigma_0}\right)^\beta
            \left(\frac{r}{a_0}\right)^\gamma
            \left(\frac{T}{T_0}\right)^\alpha\ .
\label{eq:viscosity}
\end{equation}
For small moonlets ($h \le 50$m) one has a rather small ratio $T_\text{ini}/T_0 \le 5$. The
power law dependence (\ref{eq:viscosity}) with $\alpha = 1/2$ \citep{salo01} then gives a
viscosity about twice as large as the unperturbed one. On the other hand, the surface mass
density in the gap region after the scattering by the moonlet is about half of the unperturbed
value, so that a constant viscosity $\nu_0$ can be justified.  In case of large moonlets the
ratio $T_\text{ini}/T_0$ is by far higher ($190$ for a moonlet with $300$m Hill radius at
$|\tilde{x}| = 2$). The viscosity will rise considerably with increasing temperature, tending
to an increased heating, and thus, to an increased cooling time scale. However, because of an
increasing coefficient of restitution and due to increased collision rates, an enhanced cooling
works against that effect. Insofar, as a first step, we have used a constant viscosity and a
constant coefficient of restitution, interpreting both as effective values.

Furthermore, we have neglected the temperature decrease due to heat conduction, based on the grounds that
the relaxation of the ring temperature by granular cooling is the faster process, compared to the
large diffusive time scales blurring the propeller gaps. However, heat conduction as well as viscous
diffusion could become important for small moonlets. We have estimated the temperature decrease
due to heat conduction for a simplified model, described in \ref{app:heat_conduction}. For
small moonlets ($h \le 50$m) the temperature decrease due to heat conduction is comparable to
the temperature decrease of our granular cooling driven solution (\ref{eq:tsolution}). On the
other hand, for large moonlets the temperature decrease due to heat conduction was approximatly
$20$ times smaller than our cooling driven relaxation after $10$ orbits, justifying a neglect of
heat conduction for large moonlets, and thus, our approach.


\begin{figure}
\begin{adjustwidth}{-1in}{-1in}%
\begin{subfigure}[t]{0.7\textwidth}
    \centering
    \includegraphics[width=0.9\textwidth]{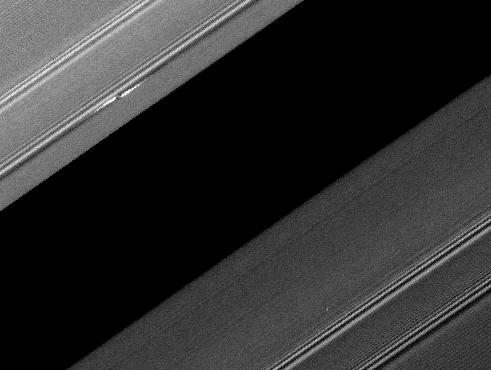}
    \caption{}
    \label{fig:earhart_before}
\end{subfigure}%
\qquad
\begin{subfigure}[t]{0.7\textwidth}
    \centering
    \includegraphics[width=0.7\textwidth]{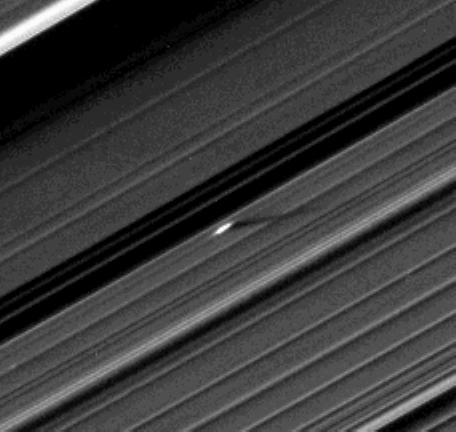}
    \caption{}
    \label{fig:earhart_equinox}
\end{subfigure}%
\caption{The propeller moonlet Earhart near the Enke gap. (a) This image was taken by the
         Cassini spacecraft on April 11, 2008 with Cassini's narrow angle camera. The resolution
         is about $2$ km per pixel. The propeller structure is about $5$ km in radial dimension
         and about $60$ km in azimuthal dimension. (b) Image of Earhart near Saturn's equinox,
         taken by Cassini's narrow angle camera on August 13, 2009. The resolution is $7$ km/pixel.
         The shadow cast by the propeller is $350$km long and the height of the propeller is
         estimated to be $260$m \citep{tiscareno10}.
         Credit: NASA/JPL/Space Science Institute}\label{fig:earhart}
\end{adjustwidth}%
\end{figure}

In the summer of 2009, near Saturn's equinox, the Cassini spacecraft took images, which show
prominent shadows cast by propeller moonlets. The height of the propeller features were calculated from
the observed shadow length \citep{tiscareno10}. For the propeller moonlet Bl\'{e}riot, a height of
$430\pm30$m was determined, for Santos-Dumont $120$m, and for Earhart $260$m. Figure \ref{fig:earhart}
shows the propeller moonlet Earhart orbiting near the Encke gap, where Figure \ref{fig:earhart_equinox}
portraits Earhart in August 2009, a few days after Saturn's equinox, to cast a $350$km long shadow.
In contrast, Figure \ref{fig:earhart_before} shows Earhart in April 2008 long before equinox.

Using our values of the gap averaged propeller height from Table \ref{table:heights}, the
height of the propeller structure is approximately $0.7$ times the Hill radius of the propeller
moonlet. This gives $370$m for the Hill radius of Earhart and $615$m for the Hill radius of Bl\'eriot.
Assuming a sperical shape, the moonlet radius of Earhart is then $240$m for a mass density of
$600\,\text{kg}/\text{m}^3$, and $280$m for a mass density of $400\,\text{kg}/\text{m}^3$.

We determined the azimuthal extent of Earhart from Figure \ref{fig:earhart_equinox} to be
$120\pm20$km, the Cassini ISS team reported about $130$km on their website. The propeller
structure thus azimuthally extends about $60\pm10$km downstream from the moonlet. In Section
\ref{sec:energy_balance} we used a ring temperature of $1.05\cdot T_\text{eq}$ to determine
the end of the exponential temperature decay. Using this criterium for a moonlet with $370$m
Hill radius and for the radial ring particle position $|\tilde{x}| = 2.5$ (approximately the
middle of the gap region), the exponential temperature decay stops after about $7$ orbits.
These $7$ orbits at $|\tilde{x}| = 2.5$ corresponds to an azimuthal propeller length of $61$km
downstream of the moonlet which agrees astonishingly well with the $60\pm10$km determined from
Figure \ref{fig:earhart_equinox}. To calculate the mean decay constant $\langle \gamma \rangle$,
we used equation (\ref{eq:ymax}) with the determined azimuthal propeller extent, to find
$\langle\gamma\rangle =  0.05$.  At $|\tilde{x}| = 2.5$ this corresponds to
$k_3 (1-\varepsilon^2) = 1.1$, which nicely compares
to $k_3 (1-\varepsilon^2) = 1.125$ used in our model.



\section{Summary and Outlook}
\label{sec:conclusions}

In the present work we study the vertical extent of propeller structures in Saturn's
rings. Our focus lies on the gap region of the propeller and on non-inclined propeller
moonlets. In order to describe the vertical structure of propellers we extend the model
of \citet{spahn00} to include the vertical direction. We use the scattering operator
concept \citep{spahn89} to model the gravitational interaction of the embedded moonlet
with the ring particles, taking place in a relatively small region around the moonlet.
Outside this region we describe the Keplerian ring flow using hydrodynamic equations.
The azimuthal relaxation of the propeller height is modelled by the disturbed balance
of viscous heating and granular cooling using the energy balance equation of the ring
particle's thermal motion.

Our conclusions and results are:

\begin{enumerate}
\item
Considering only the gravitational interaction of ring particles with a non-inclined
moonlet, the moonlet does not induce considerable vertical excursions of ring particles.
However, the moonlet causes considerable thermal motion in the ring plane.
The maximal value of this thermal motion, when measured by the radial thermal velocity,
scales well with the Hill radius according to $c_x \sim h\Omega$.
\item
Collisions between ring particles are an effective mechanism to locally convert the
lateral induced thermal motion into vertical excursions of ring particles. After the
scattering by the moonlet, the ratio of vertical to radial thermal velocity is far from
its equilibrium value, but the local relaxation to the equilibrium value takes only a few
collisions per particle \citep{haemeen80, haemeen93}. Consequently, after a few collisions
per particle, the ring temperature describes the vertical extent of the propeller structure
astonishingly well.
\item
We use half of the effective geometric thickness of the ring as a measure for the
propeller height in the gap region and find maximal heights on the order of the Hill
radius of the moonlet. The height, averaged over the gap region, is obtained to be
about $0.7$ Hill radii.
\item
For the first few orbits, the azimuthal evolution of the ring temperature is characterised
by an exponential decay to a local equilibrium temperature. Later, the viscous heating
and the collisional cooling are balanced and the ring temperature evolves as a function
of the local optical depth of the ring. The fast decay is consistent with images of Earhart,
taken by the Cassini spacecraft near Saturn's equinox. We found that for the exponential
decay phase the neglect of heat conduction is a useful and valid assumption for large
propeller moonlets like Bl\'{e}riot and Earhart. For small moonlets, e.g. with Hill
radius of $50$m, heat conduction and dissipative effects become important right after the
scattering.
\end{enumerate}

Future work should incorporate a kinetic description of the counteracting viscous heating
and collisional cooling, to study the consequences of a velocity dependent coefficient of
restitution and a temperature dependent viscosity on the azimuthal relaxation of the ring
temperature.

Another subject of ongoing work is the vertical extent of the wake region of
propellers. By dropping the averaging process in the calculation of the scattering operator,
radial and vertical phase information of the ring particle's motion is preserved. This allows
us to describe the coherent motion of ring particles in the wake region. An extension to
inclined moonlets would then allow comparison to the wake region of the ring-moon Daphnis.

Further work should be done on the inclusion of heat conduction, viscous diffusion and
fluctuations into our model, to better describe propellers of small moonlets. Finally, we
plan to compare our results to direct N-body simulations of propellers \citep{seiss05, sremcevic07}.

\section*{Acknowledgements}

We thank Heikki Salo for very useful discussions. We kindly acknowledge the efforts of the
Cassini ISS team in the design and operation of the ISS instrument. This work was supported
by Deutsche Forschungsgemeinschaft (Sp 384/24-1; 25-1).


\appendix
\section{Influence of heat conduction on the temperature decay}
\label{app:heat_conduction}

We neglected the heat conduction term in equation (\ref{eq:temperatur_simplified})
on the grounds that the relaxation of the ring temperature is, compared to the mass
diffusion, a fast process. To check for which parameters this assumption is valid,
we use our solution of (\ref{eq:temperatur_simplified}) to estimate how large a heat
conduction term in equation (\ref{eq:temperatur_further_simplified}) would have been.
With a heat conduction term, equation (\ref{eq:temperatur_further_simplified}) reads
\begin{equation}
\tilde{x}\ \frac{\partial T_1}{\partial\tilde{y}} = 
\nu_0 \Omega_0 \frac{\sigma_1}{\Sigma_0} + \frac{4}{9} k_3 \tau (1-\varepsilon^2) T_1
- \frac{4\kappa}{9\Omega_0 h^2}
      \left( \frac{\partial^2 T_1}{\partial\tilde{x}^2} +
             \frac{\partial^2 T_1}{\partial\tilde{y}^2} \right)\ ,
\label{eq:temp_with_heat}
\end{equation}
with the kinematic heat conductivity $\kappa$, which we assumed to be constant. Further
we assume $\kappa/\nu \approx 20$ \citep{salo01}.
The azimuthal part of the heat conduction term is several times smaller than the
cooling term which describes the inelastic collision between ring particles. For
a moonlet with $50$m Hill radius, it is at least $20$ times smaller, whereas it
is at least two orders of magnitude smaller for a moonlet with $200$m Hill radius.
The radial part of the heat conduction term, on the other hand, is of the order of
the cooling term for $|\tilde{x}| < 1.5$. For small moonlets, e.g. $h=50$m, this applies
also for $|\tilde{x}| < 4$.

On the timescale of the mass diffusion process, heat conduction can not be neglected,
but for small moonlets heat conduction seems to be also important in the first few orbits.
Therefore, we estimated the temperature decrease in the middle of the gap at $\tilde{x} = -2.5$.
For that, we assumed a Gaussian radial temperature profile at $\tilde{y} = 0^+$, centered at
$\tilde{x}_0 = -2.5$ and with standard deviation $\sigma$ calculated according to full width
at half maximum of $3$:
\begin{equation}
T(\tilde{x}, 0) = T(\tilde{x}_0, 0) \exp\left(
   - \frac{(\tilde{x}-\tilde{x}_0)^2}{2\sigma^2} \right)\ .
\end{equation}
Using this as initial condition, we solved equation (\ref{eq:temp_with_heat}) neglecting
all but the heat conduction term
\begin{equation}
\tilde{x}_0\ \frac{\partial T_1}{\partial\tilde{y}} = 
- \frac{4\kappa}{9\Omega_0 h^2}
      \left( \frac{\partial^2 T_1}{\partial\tilde{x}^2} +
             \frac{\partial^2 T_1}{\partial\tilde{y}^2} \right)\ ,
\label{eq:temp_only_heat}
\end{equation}
further simplified by setting $\tilde{x} = \tilde{x}_0$. The azimuthal evolution of the
temperature at radial position $\tilde{x}_0$ ist then given by
\begin{equation}
T(\tilde{x}_0, \tilde{y}) = \sqrt{2\pi\sigma^2}\,T(\tilde{x}_0,0)\,
   \left( 2\pi\sigma^2 + \frac{16\pi \kappa}{9\Omega_0 h^2\tilde{x}_0} \tilde{y} \right)^{-1/2}\ .
\end{equation}
For a moonlet with a Hill radius of $50$m, the temperature decrease according to the above
solution was approximatly equal to the temperature decrease of our solution
(\ref{eq:tsolution}) in the first few orbits. For a moonlet with $200$m Hill radius on the
other hand, the temperature decrease due to heat conduction after $10$ orbits was approximatly
$20$ times smaller than the temperature decrease of our solution (\ref{eq:tsolution}). This
indicates, although we neglected the Kepler shear by setting $\tilde{x} = \tilde{x}_0$ in
equation (\ref{eq:temp_only_heat}), the importance of heat conduction for the temperature decay
for small moonlets. For large moonlets, $200$m Hill radius and above, the neglect of heat
conduction seems to be a valid assumption.

\bibliographystyle{model2-names}

\begin{thebibliography}{00}



\bibitem[Araki and Tremaine(1986)]{araki86}
Araki, S., Tremaine, S., 1986,
The Dynamics of Dense Particle Disks,
Icarus, vol. 65, pp.83-109.

\bibitem[Artymowicz(2006)]{artymowicz06}
Artymowicz, P., 2006,
Planetary Systems,
Am. Inst.  Physics Conf. Series, vol. 843, pp.3-34.


\bibitem[Borderies et al.(1985)]{borderies85}
Borderies, N., Goldreich, P., Tremaine, S., 1985,
A Granular Flow Model for Dense Planetary Rings,
Icarus, vol. 63, pp.406-420.

\bibitem[Burns and Cuzzi(2006)]{burns06}
Burns, J.A.  and Cuzzi, J.N.C., 2006,
Our local astrophysical laboratory,
Science, vol. 312, pp.1753-1755.

\bibitem[Goldreich and Tremaine(1978)]{goldreich78}
Goldreich, P. and Tremaine, S., 1978,
Saturn's rings velocity dispersion,
Icarus, vol. 34, pp.227-239.

\bibitem[H\"{a}meen-Anttila and Lukkari(1980)]{haemeen80}
H\"{a}meen-Anttila, K. A. and Lukkari, J., 1980,
Numerical simulations of collisions in Keplerian systems,
A\&AA, vol. 71, pp. 475-497.

\bibitem[H\"{a}meen-Anttila and Salo(1993)]{haemeen93}
H\"{a}meen-Anttila, K. A. and Salo, H., 1993,
Generalized Theory of Impacts in Particulate Systems,
EM\&P, vol. 62, pp. 47-84.

\bibitem[H\'{e}non and Petit(1986)]{henon86}
H\'{e}non, M. and Petit, J.-M., 1986,
Series expansion for encounter-type solutions of Hill's problem,
Celestial Mechanics, vol. 38, pp. 67-100.

\bibitem[Hill(1878)]{hill78}
Hill, G.W., 1878,
Researches in the Lunar Theory,
Am. J. Math, vol. 1, pp. 5-26, 129-147, 245-260.

\bibitem[Ida and Makino(1992)]{ida92}
Ida, S., Makino, J., 1992,
N-body simulation of gravitational interaction between planetesimals and a protoplanet. I - Velocity distribution of planetesimals,
Icarus, vol. 96, pp. 107-120.

\bibitem[Jacobson et al.(2008)]{jacobson08}
Jacobson R. A. et al., 2008,
Revised orbits of Saturn's small inner satellites,
AJ, vol. 135, pp. 261-263.

\bibitem[Lewis and Stewart(2000)]{lewis00}
Lewis, M.~C., Stewart, G.~R., 2000,
Collisional dynamics of perturbed planetary rings I,
AJ, vol. 120, pp.3295-3310.

\bibitem[Lewis and Stewart(2009)]{lewis09}
Lewis, M.~C., Stewart, G.~R., 2009,
Features around embedded moonlets in Saturn's rings: The role of self-gravity and particle size distributions,
Icarus, vol. 199, pp.387-412.

\bibitem[Lissauer(1993)]{lissauer93}
Lissauer, J.J., 1993,
Planet formation,
ARA\&A, vol. 31, pp. 129-174.

\bibitem[Ohtsuki and Emori(2000)]{ohtsuki00}
Ohtsuki, K., Emori, H., 2000,
Local N-Body Simulations for the Distribution and Evolution of Particle Velocities in Planetary Rings,
AJ, vol. 119, pp. 403-416.

\bibitem[Papaloizou et al.(2007)]{papaloizou07}
Papaloizou, J.C.B. et al.  2007,
Disk-Planet Interactions During Planet Formation,
Protostars and Planets V, pp. 655-668.

\bibitem[Petit and H\'{e}non(1986)]{petit86}
Petit, J.-M., H\'{e}non, M., 1986,
Satellite encounters,
Icarus, vol. 66, pp. 536-555.

\bibitem[Petit and H\'{e}non(1987)]{petit87}
Petit, J.-M., H\'{e}non, M., 1987,
A numerical simulation of planetary rings. I -- Binary encounters,
A\&A, vol. 173, pp. 389-404.

\bibitem[Press et al.(1992)]{press92}
Press, W. H., Teukolsky, S. A., Vetterling, W. T., Flannery, B. P., 1992,
Numerical Recipes in C (2nd ed.),
Cambridge University Press.


\bibitem[Salo(1991)]{salo91}
Salo, H., 1991,
Numerical Simulations of Dense Collisional Systems,
Icarus, vol. 90, pp. 254-270.

\bibitem[Salo et al.(2001)]{salo01}
Salo, H., Schmidt, J., Spahn, F., 2001,
Viscous Overstability in Saturn's B Ring. I. Direct Simulations and Measurement of Transport Coefficients,
Icarus, vol. 153, pp. 295-315.

\bibitem[Schmidt et al.(1999)]{schmidt99}
Schmidt, J., Salo, H., Petzschmann, O., Spahn, F., 1999,
Vertical distribution of temperature and density in a planetary ring,
A\&A, vol. 345, pp. 646-652.

\bibitem[Sei{\ss } et al.(2005)]{seiss05}
Sei{\ss }, M., Spahn, F., Srem{\v c}evi{\'c}, M., Salo, H., 2005,
Structures induced by small moonlets in Satrun's rings: Implications for the Cassini mission,
Geophys. Res. Lett., vol. 32, doi:10.1029/2005GL022506.

\bibitem[Showalter et al.(1986)]{showalter86}
Showalter, M.~.R., Cuzzi, J.~N., Marouf, E.~A., Esposito, L.~W., 1986,
Satellite ``Wakes'' and the Orbit of the Encke Gap Moonlet,
Icarus, vol. 66, pp. 297-323.
    
\bibitem[Simon and Jenkins(1994)]{simon94}
Simon, V., Jenkins, J.~T., 1994,
On the Vertical Structure of Dilute Planetary Rings,
Icarus, vol. 110, pp. 109-116.

\bibitem[Spahn(1987)]{spahn87}
Spahn, F., 1987,
Scattering Properties of a Moonlet (Satellite) Embedded in a Particle Ring:
Application to the Rings of Saturn,
Icarus, vol. 71, pp. 69-77.

\bibitem[Spahn and Wiebicke(1989)]{spahn89}
Spahn, F., Wiebicke, H. J., 1989,
Long-Term Gravitational Influence of Moonlets in Planetary Rings,
Icarus, vol. 77, pp. 124-134.

\bibitem[Spahn, Scholl and Hertzsch(1994)]{spahn94}
Spahn, F., Scholl, H., Hertzsch, J.-M., 1994,
Structures in planetary rings caused by embedded moonlets,
Icarus, vol. 111, pp. 514-535.

\bibitem[Spahn and Srem\v{c}evi\'c(2000)]{spahn00}
Spahn, F., Srem\v{c}evi\'c, M., 2000,
Density patterns induced by small moonlets in Saturn's rings?,
A\&A, vol. 358, pp. 368-372.

\bibitem[Spahn et al.(2000)]{spahn00b}
Spahn, F., Schmidt, J., Petzschmann, O., Salo, H., 2000,
Note: Stability analysis of a Keplerian disk of granular grains: Influence of thermal diffusion,
Icarus, vol. 145, pp. 657-660.

\bibitem[Srem\v{c}evi\'{c}, Spahn, and Duschl(2002)]{sremcevic02}
Srem\v{c}evi\'{c}, M., Spahn, F., Duschl, W.J., 2002,
Density structures in perturbed thin cold discs,
Mon. Not. R. Astrom. Soc., vol. 337, 1139-1152.
    
\bibitem[Srem\v{c}evi\'{c} et al.(2007)]{sremcevic07}
Srem\v{c}evi\'{c}, M., Schmidt, J., Salo, H., Sei{\ss}, M., Spahn, F., Albers, N., 2007,
A belt of moonlets in Saturn's A ring,
Nature, vol. 449, 1019-1021.
    
\bibitem[Stewart et al.(1984)]{stewart84}
Stewart, G. R., Lin, D. N. C., Bodenheimer, P., 1984,
Collision-induced transport processes in planetary rings.
In Planetary Rings (R. Greenberg and A. Brahic, Eds.),
pp. 447-512, Univ. of Arizona Press, Tucson.

\bibitem[Tiscareno et al.(2006)]{tiscareno06}
Tiscareno, M.~S., Burns, J.~A., Hedman, M.~M., Porco, C.~C., Weiss, J.~W.,
  Murray, C.~D., and Dones, L., 2006,
Observation of ``propellers'' indicates 100-metre diameter moonlets reside in Saturn's A-ring.
\newblock {\em Nature}, vol. 440, pp. 648-652.

\bibitem[Tiscareno et al.(2007)]{tiscareno07}
Tiscareno, M. S. et al., 2007,
Cassini imaging of Saturn's rings. II. A wavelet technique for analysis of density waves and other radial structure in the rings,
Icarus, vol. 189, pp. 14-34.

\bibitem[Tiscareno et al.(2008)]{tiscareno08}
Tiscareno, M.~S., Burns, J.~A., Hedman, M.~M.; Porco, C.~C., 2008,
The Population of Propellers in Saturn's A Ring,
Astronomical Journal, vol. 135, pp. 1083-1091.

\bibitem[Tiscareno et al.(2010)]{tiscareno10}
Tiscareno M. S. et al., 2010,
Physical Characteristics and Non-Keplerian Orbital Motion of ``Propeller'' Moons Embedded in Saturn's Rings,
ApJL, vol. 718, pp. L92-L96.

\bibitem[Weiss et al.(2009)]{weiss09}
Weiss, J. W., Porco, C. C., Tiscareno M. S., 2009,
Ring Edge Waves and the Masses of Nearby Satellites,
Astronomical Journal, vol. 138, pp. 272-286.

\end{thebibliography}

\end{document}